\documentclass[journal=jacsat,manuscript=article]{achemso}

\usepackage[version=3]{mhchem} % Formula subscripts using \ce{}
\usepackage{lineno}
\usepackage{booktabs}
\usepackage{physics}
\usepackage{bm}
\usepackage{subcaption}

\newcommand\Rey{\mbox{\text{Re}}}  % Reynolds number

\author{Ashray Mohit}
\affiliation{Department of Mechanical Engineering, University of Michigan, Ann Arbor, USA}
\email{ashraym@umich.edu}
\phone{+1(734)-834-6960}
\author{Jenna Stolzman}
\affiliation{Department of Mechanical Engineering, University of Michigan, Ann Arbor, USA}

\author{Margaret Wooldridge}
\affiliation{Department of Mechanical Engineering, University of Michigan, Ann Arbor, USA}
\alsoaffiliation{Department of Aerospace Engineering, University of Michigan, Ann Arbor, USA}

\author{Jesse Capecelatro}
\affiliation{Department of Mechanical Engineering, University of Michigan, Ann Arbor, USA}
\alsoaffiliation{Department of Aerospace Engineering, University of Michigan, Ann Arbor, USA}

\title[An \textsf{achemso} demo]
  {Effect of Nozzle Geometry on the Performance of Non-Assist Flares}

\begin{document}

% \begin{tocentry}

% Some journals require a graphical entry for the Table of Contents.
% This should be laid out ``print ready'' so that the sizing of the

% Inside the \texttt{tocentry} environment, the font used is Helvetica
% 8\,pt, as required by \emph{Journal of the American Chemical

% The surrounding frame is 9\,cm by 3.5\,cm, which is the maximum
% permitted for  \emph{Journal of the American Chemical Society}
% graphical table of content entries. The box will not resize if the
% content is too big: instead it will overflow the edge of the box.

% This box and the associated title will always be printed on a
% separate page at the end of the document.

% \end{tocentry}


\begin{abstract}
 This study employs large-eddy simulations with a flamelet progress variable approach to systematically quantify the influence of nozzle geometry on combustion efficiency, mixing, and blowout resistance in non-assist methane flares. Five canonical nozzle shapes--circle, low aspect ratio ellipse, high aspect ratio ellipse, diamond, and square--were evaluated under relevant industrial flare conditions. Results demonstrate that cornered geometries enhance near-field recirculation, promote mixing, and sustain flame attachment, resulting in up to a $5\%$ improvement in combustion efficiency compared with streamlined nozzles. The square nozzle performed best irrespective of the wind direction (orientation) and maintained a combustion efficiency $> 96.5\%$ even at the highest tested crosswind velocities, while other streamlined designs exhibited early flame lift-off, reduced recirculation, and efficiency losses. Analysis of mixing and vorticity reveals that sharp-edged nozzles accelerate scalar homogenization and buffer flames against crosswind-induced strain, directly translating to increased blowout resistance. 
\end{abstract}

\section{Introduction}
Waste gas flares play a crucial role in industrial settings by safely combusting surplus gases that cannot be economically recovered or utilized. These systems are typically classified as air-assist, steam-assist, pressure-assist, or non-assist flares. Non-assist flares operate without supplemental air or steam injection, relying exclusively on the inherent momentum of the waste gas and ambient air to promote mixing and combustion. The United States Environmental Protection Agency (EPA) mandates that non-assist flares achieve combustion efficiencies greater than 96.5\% to adequately limit emissions and minimize environmental harm\cite{epa2025code}. Achieving and maintaining such high efficiency, however, becomes increasingly difficult under strong crosswind conditions.

The presence of crosswinds introduces complex aerodynamic effects that can deform flame shapes, destabilize the flame front, and alter mixing processes between fuel and air. Prior experimental and numerical investigations have established that crosswinds can degrade flame stability and lower combustion efficiency \cite{gogolek2002efficiency,johnson2002parametric,burtt2014efficiency,stolzman2025experimental,stolzman2025assesing,stolzman2024effects}. To address these challenges, various engineered flare tip designs have been developed, incorporating features such as air-assist jets, steam injection, or specialized geometries to enhance mixing and flame stability \cite{allen2011tceq,gogolek2010emissions,Reed2004FlareTip}. While these engineered solutions can be effective in improving combustion efficiency and ensuring compliance under adverse conditions, they often entail higher installation and operational costs. As a result, most oil and gas facilities continue to favor the use of simple round utility pipes as flare tips, owing to their ease of integration and minimal capital expenditure \cite{stolzman2025shrouds}. This preference for standard non-assist flare configurations persists despite their susceptibility to efficiency losses in crosswind environments, underscoring the importance of understanding and optimizing their performance within these practical constraints.

Gas flares are often modeled as jets in crossflow (JICF), where an injected jet is introduced into a moving stream of air that is perpendicular or at an angle to the jet direction \cite{gaipl2025combustion,mohit2025impact}. They have been the subject of extensive investigations, as they offer a representative platform for understanding mixing, combustion, and pollutant formation. Research in this area has provided valuable insights into jet deflection, entrainment and mixing processes, the formation of characteristic vortex patterns, and the mechanisms that govern flame stabilization when jets interact with ambient flows \cite{roshko1976structure,smith1998mixing,mahesh2013interaction,fric1994vortical,gruber2018direct,pitsch2006large}.

The canonical JICF configuration typically employs a flush-mounted circular nozzle issuing into a crossflow, serving as a well-established baseline for experimental and numerical studies.
Previous studies have demonstrated that the nozzle shape plays a significant role in dictating downstream flow characteristics, including jet penetration, mixing efficiency, and the formation of vortex structures \cite{haven1997kidney,liscinsky1996crossflow}. Salewski et al. \cite{salewski2008mixing} found that non-circular nozzles promote enhanced turbulent mixing and more complex vortex interactions compared with their circular counterparts, due to interactions of the secondary flow structures with the jet core. 

The nozzle shape has also been found to affect combustion performance. Compared with circular nozzles, high-aspect-ratio elliptical nozzles oriented perpendicular to the crossflow generate stronger secondary streamwise vortices downstream of the primary counter-rotating vortex pair (CVP), significantly accelerating turbulent mixing between fuel and air \cite{new2003elliptic}. This enhanced mixing improves flame stability under lean conditions in gas turbine combustors, reducing emissions like soot and NOx, and enables smaller combustor designs \cite{kolb2016influence}.
% In internal combustion engines, optimized nozzle shapes improve spray penetration, control droplet size, enhance air entrainment during the short mixing time, and reduce wall wetting or rich zones, directly boosting combustion efficiency and lowering particulate emissions \cite{kourmatzis2015air}.

In the context of waste gas flaring, most experimental and numerical studies have identified the jet-to-crossflow momentum flux ratio, the degree of dilution of the combustion gases, and the heating value of the flare gas as key factors controlling combustion efficiency and operational performance, demonstrating that higher momentum flux ratios and greater heating values generally promote more stable combustion and reduced pollutant emissions, whereas increased dilution tends to lower flame temperatures and reduce combustion efficiency \cite{wang2024computational,johnson2000efficiencies,stolzman2025effects}. Gaipl et al. \cite{gaipl2025combustion} showed that elevated crosswind velocities cause the flame to lift off the nozzle, allowing unburned fuel to escape into high-strain regions downstream, which decreases overall combustion efficiency.

Waste gas flares may be exposed to strong and highly variable crosswinds, and under such adverse conditions, the flame can be completely blown off the tip of the stack. When blowout occurs, the flare transitions from controlled combustion to direct venting of unburned gases, resulting in elevated emissions of methane and other pollutants. To mitigate this risk, regulations require that waste gas flares be equipped with a continuously burning pilot\cite{epa2025code}, which helps stabilize the main flame and reduce the likelihood of blowout. However, even with a pilot present, strong or highly turbulent wind gusts can still cause temporary flame blowout and subsequent venting of unburned gases. For the purposes of the present study, a continuously burning pilot flame is not considered. As such, the analysis represents a scenario in which the risk of blowout and direct venting is not mitigated by pilot stabilization. 

Numerous studies have investigated the underlying mechanisms responsible for flame blow-out in jet and bluff-body burners, highlighting the critical roles of local flow velocity, quenching scalar dissipation rates, and flame stretch in destabilizing the flame front and triggering blow-out, particularly in scenarios with high crosswind velocities and low jet momentum \cite{huang1994stability,zhu2025effect,han2000observations,wu2006blowout,kolla2012mechanisms}. One common approach to avoid flame blow-out is to introduce swirl to the jet, which imparts additional angular momentum, promotes internal recirculation zones, and anchors the flame by increasing the residence time in the high-temperature reaction zone \cite{feikema1990enhancement}. Other studies have explored the use of flame holders or pilot flames as stabilization mechanisms. Flame holders can shield the flame from crosswind disturbances, while pilot flames help sustain combustion under challenging flow conditions, both working to delay the onset of blowout \cite{aniello2023influence,chaparro2006blowoff,shanbhogue2009lean}.

Despite extensive research on reacting JICF in the context of both fundamental and applied research, there remains a critical gap in systematically connecting the effects of mixing, combustion efficiency, and blowout resistance with nozzle geometry under realistic crosswind conditions. Most previous studies have treated these performance metrics in isolation, with limited consideration of how simple geometric changes to the flare tip can passively enhance performance. The present work aims to address this gap by employing high-fidelity large-eddy simulations to quantify the role of nozzle-induced flow structures on near-field mixing, combustion efficiency, and blowout dynamics in non-assist methane flares. Circular, square, and elliptical nozzle geometries are considered under varying crosswind directions, with methane flow rates up to 50,000 standard cubic feet per day (SCFD), representative of emissions from storage tanks,  \cite{torres2012emissions}  and other low-pressure/low-flow waste gas sources, and crosswind velocities reaching up to 27 m/s ($\approx$ 60 miles per hour).

The paper is organized as follows: Section 2.1\ref{subsec:2} presents the problem set-up and key metrics considered. The governing equations, discretization, methods, and validation with experimental flare studies are discussed in Section 2.2\ref{subsec:3} and Section 2.3\ref{subsec:4}. Section 3.1\ref{subsec:5} highlights the effect of nozzle geometry on the flame structure and attachment. The effect of nozzle shape on combustion efficiency and mixing is reported in Sections 3.2\ref{subsec:6} and 3.3\ref{subsec:7}, respectively. Finally, Section 3.4\ref{subsec:8} compares the blow-out velocities of the different nozzle designs.

\section{Methodology} \label{sec2}
\subsection{System description} \label{subsec:2}
\begin{figure}[h!]
    \centering
    \includegraphics[width=4.5in]{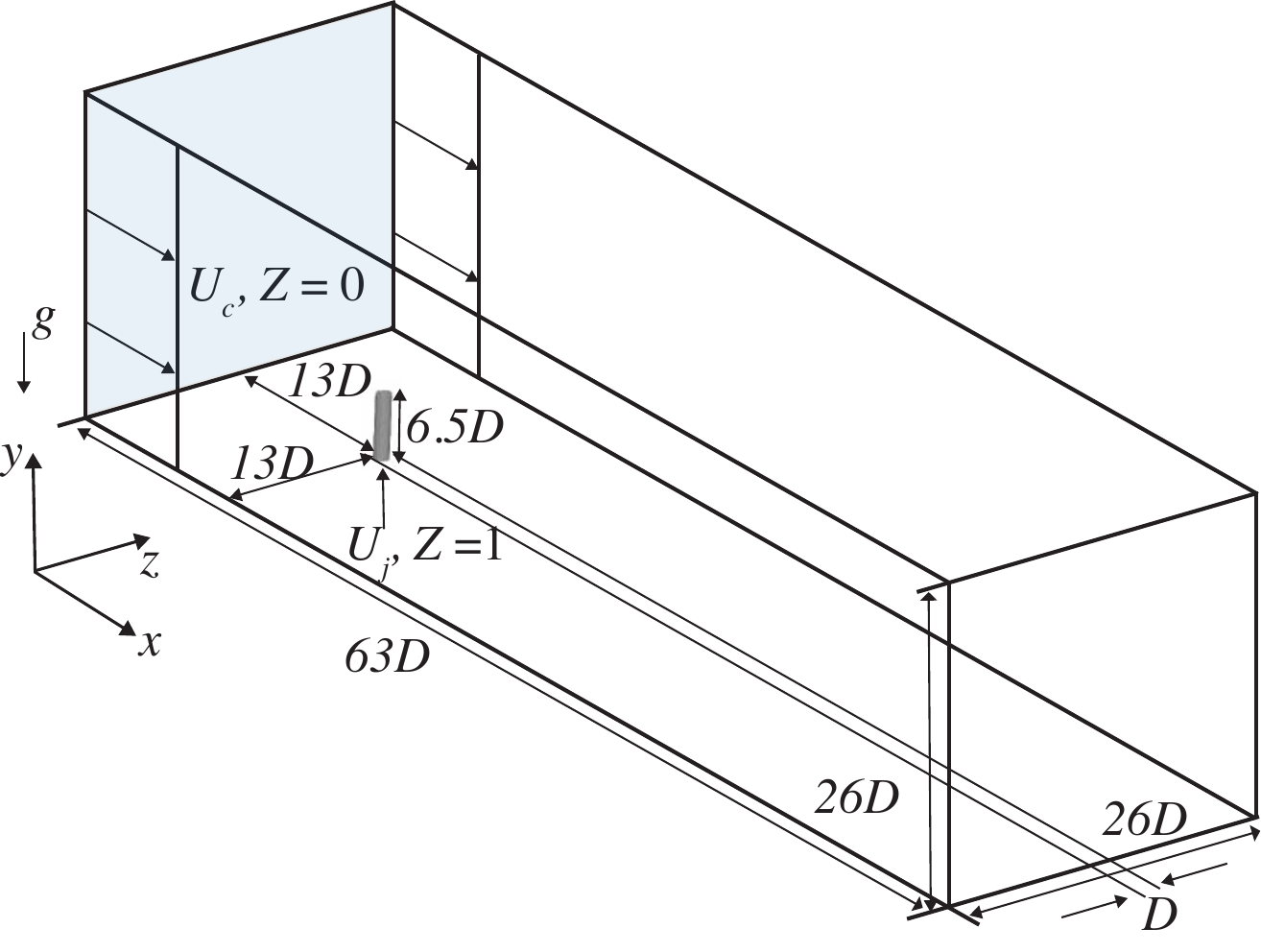}
    \caption{Schematic of the reacting jet-in-crossflow configuration. The white planes represent the domain boundaries where combustion efficiency is measured. The blue plane represents the crossflow inlet.}
    \label{fig:setup}
\end{figure}

The jet-in-crossflow (JICF) configuration, depicted in Fig.~\ref{fig:setup}, is selected as a representative model for non-assist flares. The computational domain spans $63D$ in the streamwise ($x$) direction and $26D$ in both height ($y$) and spanwise ($z$) direction. The nozzle is located $13D$ from the inlet and protrudes $6.5D$ into the domain. The nozzle location is defined as $x/D = 0$ throughout this study. These dimensions are selected to ensure that both the flame and associated vortical structures remain sufficiently far from the boundaries (shown in white), thereby minimizing the influence of the domain boundaries on the combustion dynamics.

In contrast with conventional JICF configurations, which feature a nozzle flush with the bottom wall, the present setup incorporates a nozzle that protrudes into the domain (i.e., is raised relative to the domain boundary) to capture the near-nozzle flow relevant to flare designs. The protrusion primarily results in increased jet penetration; however, the downstream trajectory and wake are still predominantly controlled by the crossflow momentum, as observed in flush-mounted arrangements. Notably, the formation and characteristics of the counter-rotating vortex pair and shear layers are largely unaffected by the nozzle extension \cite{margason1993fifty,shi2023influence}.

Fuel enters the domain with a velocity $U_j$ and a mixture fraction $Z=1$ (i.e., pure methane), while the crossflow at the inlet boundary is prescribed with velocity $U_c$ and mixture fraction of $Z=0$ (i.e., pure air). Here,
% \begin{equation}
% Z = \frac{\beta - \beta_{\text{air}}}{\beta_{\text{CH}_4} - \beta_{\text{air}}},
% \end{equation}
% where $\beta = 2Y_{\rm C}/W_{\rm C} + 0.5Y_{\rm H}/W_{\rm H} - Y_{\rm O}/W_{\rm O}$ is a conserved scalar. Here, $Y_{\rm H}$, $Y_{\rm O}$, and $Y_{\rm C}$ represent the mass fractions of hydrogen, oxygen, and carbon, respectively, while $W_{\rm H}$, $W_{\rm O}$, and $W_{\rm C}$ denote their respective atomic weights. The values for the conserved scalar are $\beta_{\rm air} = -0.015$ for pure air and $\beta_{\rm CH_4} = 0.25$ for pure methane. By this definition,
$Z = 1$ indicates pure methane, and $Z = 0$ corresponds to pure air. 

The nozzle geometries considered in this study are characterized by their streamwise and spanwise dimensions, hydraulic diameter, and aspect ratio. The hydraulic diameter is defined as
\begin{equation}
    D_h = \frac{4A}{P},
\end{equation}
where $A$ is the cross-sectional area of the nozzle and $P$ is the wetted perimeter. The aspect ratio is specified as the ratio of the spanwise to the streamwise dimension for each nozzle shape.

The flow is characterized by the velocity ratio, $r = U_c / U_j$; and the jet Reynolds number, $\mathrm{Re}_j = \rho_j U_j D_h / \mu_j$, where $\rho_j$ and $\mu_j$ are the density and viscosity of the jet, respectively. The simulation parameters employed in this study are summarized in Table~\ref{tbl:Simulation_Params}.

\begin{figure}[h!]
    \centering
    \includegraphics[width=3.2in]{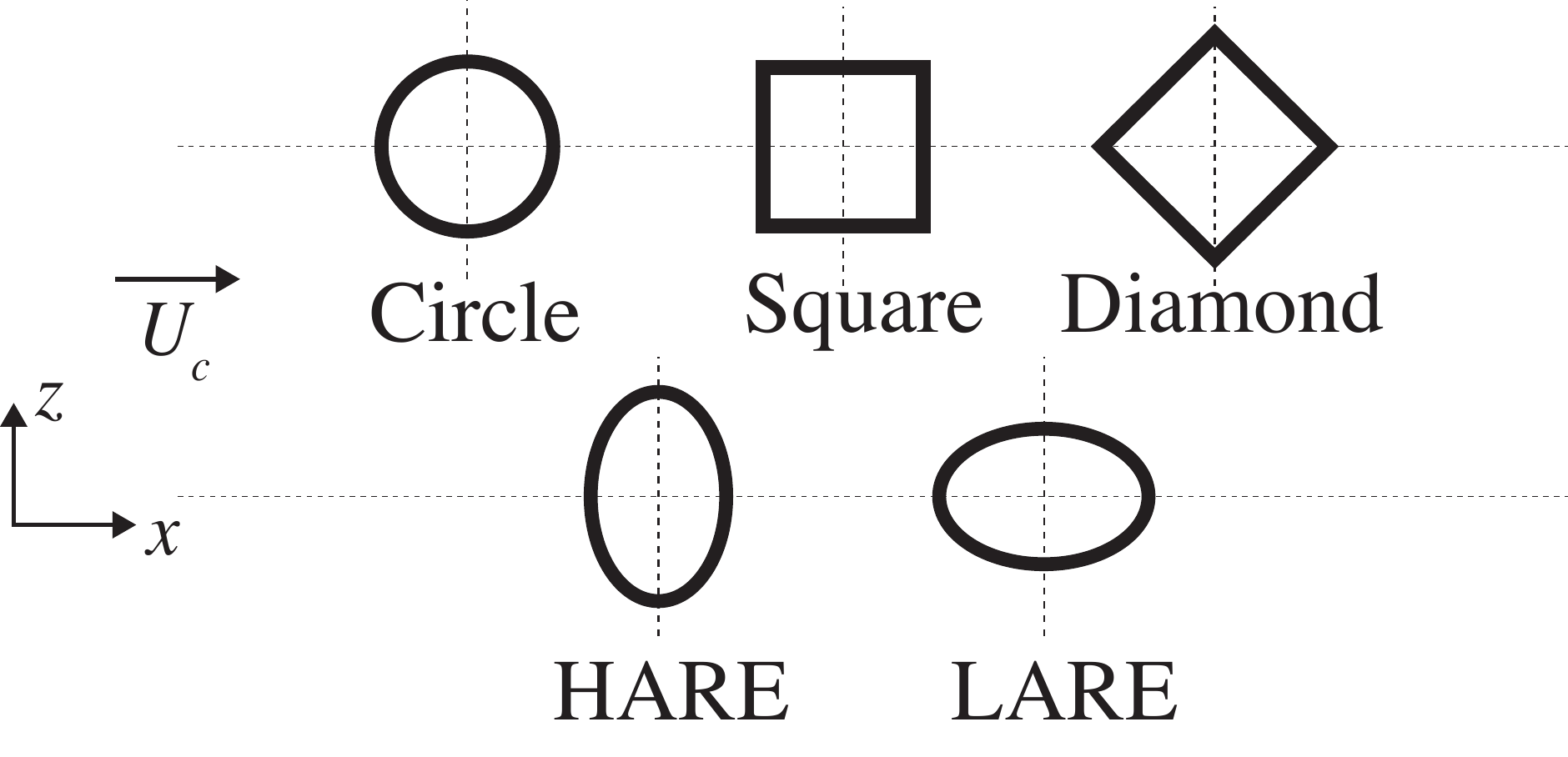}
    \caption{Cross-sectional view of the nozzle geometries considered in this study relative to the crossflow direction $U_c$.}
    \label{fig:6}
\end{figure}

The nozzle geomteries considered in this study are shown in Fig.~\ref{fig:6} and their dimensions are summarized in Table~\ref{tab:2}. Similar shapes have also been considered by Salewski et al.\cite{salewski2008mixing}, Liscinsky et al.\cite{liscinsky1996crossflow}, and Haven et al.\cite{haven1997kidney} to study the downstream mixing in a JICF. Each nozzle has the same cross-sectional area. The elliptic nozzle with the major axis parallel to the streamwise direction is referred to as a low-aspect ratio ellipse (LARE) and the elliptic nozzle with the major axis perpendicular to the streamwise direction is referred to as a high-aspect ratio ellipse (HARE). The diamond nozzle refers to a square nozzle oriented at a 45$^{\circ}$ angle with respect to the crosswind. All normalizations with $D$ are based on the diameter of the circular nozzle.

\begin{table}
  \caption{Simulation parameters considered in this study.}
  \label{tbl:Simulation_Params}
  \begin{tabular}{lllllll}
    \hline
    Shape & $U_c$ [m/s] & $U_j$ [m/s] & $r~[-]$ & \Rey$_j ~[-]$   \\
    \hline
    Circle   &  $0-24.4$ & $3.4$ & $0-7.17$  & $15451$ \\
    HARE & $0-26.8$ & $3.4$ & $0-7.88$  & $15077$ \\
    LARE & $0-23.6$ & $3.4$ & $0-6.94$  & $15077$\\
    Square & $0-26.4$ & $3.4$ & $0-7.76$  & $13687$\\
    Diamond  & $0-25.6$ & $3.4$ & $0-7.52$ & $19364$ \\
    \hline
  \end{tabular}
\end{table}

\begin{table}
\caption{Nozzle dimensions considered in this study.}
\label{tbl:Nozzle_Dims}
\begin{tabular}{lllll}     
        \hline
        {Nozzle shape} & \multicolumn{2}{c}{{Dimensions [cm]}} & Hydraulic Diameter [cm] & Aspect Ratio \\
        \cmidrule(lr){2-3}
        & {Streamwise} & {Spanwise} \\
        \hline
        Circle         & 7.62    & 7.62 & 7.62 & 1  \\
        HARE  & 6.1 & 9.52 & 7.44 & 1.56  \\
        LARE & 9.52    & 6.1 & 7.44 & 0.64  \\
        Square           & 6.75 & 6.75 & 6.75 & 1 \\
        Diamond   & 9.55 & 9.55 & 9.55 & 1\\
        \hline
    \end{tabular}
            \label{tab:2}
\end{table}

%%%%%%%%%%%%%%%%%%%%%%%%%%%%%%%%%%%%%%%%%%%%%%%%%%%%%%%%%%%%%%%%%%%%%%%%%%%%%%%
\subsection{Governing equations and numerics}\label{subsec:3}
Large-eddy simulations (LES) are conducted using the custom-developed solver \texttt{umFlameletFoam} \cite{hassanaly2018minimally}, implemented on the \texttt{OpenFOAM} platform. The solver utilizes a low-Mach-number, variable-density formulation. The governing equations for mass and momentum conservation are expressed as
\begin{equation}
\frac{\partial \rho}{\partial t} + \nabla \cdot (\rho \bm{u}) = 0,
\end{equation}
and
\begin{equation}
\frac{\partial (\rho \bm{u})}{\partial t} + \nabla \cdot (\rho \bm{u} \bm{u}) = -\nabla p + \nabla \cdot \left[ (\mu + \mu_t) \left( \nabla \bm{u} + (\nabla \bm{u})^{\sf T} - \frac{1}{3}(\nabla \cdot \bm{u}) \bm{I} \right) \right] + \rho \bm{g},
\end{equation}
where $\rho$ is the fluid density, $\bm{u}$ is the velocity vector, $p$ is pressure, $\bm{g}$ denotes gravitational acceleration, and $\bm{I}$ is the identity tensor. The turbulent viscosity, $\mu_t$, is evaluated using the dynamic Smagorinsky model proposed by Lilly \cite{lilly1992proposed}.

Combustion chemistry is modeled via the flamelet progress variable approach (FPVA) \cite{pierce2004progress}, which leverages precomputed flamelet libraries where thermochemical properties are parameterized by the local mixture fraction $Z$ and progress variable $C$. For methane-air combustion, the progress variable is defined as $C = Y_{\rm CO_2} + Y_{\rm H_2O} + Y_{\rm CO} + Y_{\rm H_2}$, where $Y_{\rm CO_2}, Y_{\rm H_2O}, Y_{\rm CO},$ and $Y_{\rm H_2}$ denote the respective product species mass fractions.

The scalar transport equations for $Z$ and $C$ are given by
\begin{equation}
\frac{\partial (\rho Z)}{\partial t} + \nabla \cdot (\rho \bm{u} Z) = \nabla \cdot (\rho \alpha \nabla Z),
\end{equation}
and
\begin{equation}
\frac{\partial (\rho C)}{\partial t} + \nabla \cdot (\rho \bm{u} C) = \nabla \cdot (\rho \alpha \nabla C) + \rho \omega_C,
\end{equation}
where $\omega_C$ is the chemical source term for the progress variable and $\alpha$ represents the scalar diffusivity. The mixture fraction variance is computed as $\rho \widetilde{Z''^2} = C_z \rho \Delta^2 |\nabla Z|^2$, where $\Delta$ is the local grid scale and $C_z$ is computed using a dynamic approach. The mixture fraction $Z$ and its variance $\widetilde{Z''^2}$ are used to construct a probability density function (PDF) for subgrid-scale fluctuations. This, combined with a PDF conditioned on $Z$ and $C$,  $\tilde{P}(Z|C)$ (approximated as a delta function), enables reconstruction of thermochemical properties. Flamelet libraries for FPVA are generated using detailed chemistry calculations with FlameMaster \cite{pitsch1998flamemaster}, incorporating the GRI-Mech 3.0 kinetic mechanism for methane-air combustion \cite{smith1999grimech}.

Dirichlet boundary conditions are enforced at the jet and crosswind inlets,  with the jet inlet set to $Z=1$, $C=0$, and the crosswind inlet set to $Z=0$, $C=0$. Pressure at both inlets employs a zero-gradient (Neumann) condition. We impose a no-slip condition for velocity at the nozzle wall, which has a thickness of $0.1D$, while scalar quantities ($Z$, $C$) and pressure are assigned zero-gradient conditions. The remaining domain boundaries---bottom, top, right, and lateral sides---are treated as outlets, applying a clipped Neumann condition for velocity and zero-gradient conditions for all other variables.

The spatial discretization is performed on a collocated, unstructured mesh with second-order accuracy. A fractional step method is used for time advancement that separates the pressure and velocity updates to maintain kinetic energy conservation. The convective fluxes in the momentum equation are solved in a skew-symmetric form, thereby minimizing artificial numerical dissipation \cite{ham2004energy}. The computational mesh employs a fully unstructured arrangement of tetrahedral cells. Each simulation utilizes approximately 12 million cells, with local grid spacing $\Delta$ from $D/28 \leq \Delta \leq D/17$. The average grid spacing, $\bar{\Delta}$, is $D/19$, consistent with the grid refinement analysis presented in our previous work~\cite{mohit2025impact}. All statistics are time averaged for $\tau=450D/U_j$ to ensure convergence unless mentioned otherwise. 

\subsection{Validation}\label{subsec:4}

\begin{figure}[h!]
    \centering
    \includegraphics[width=6.5in]{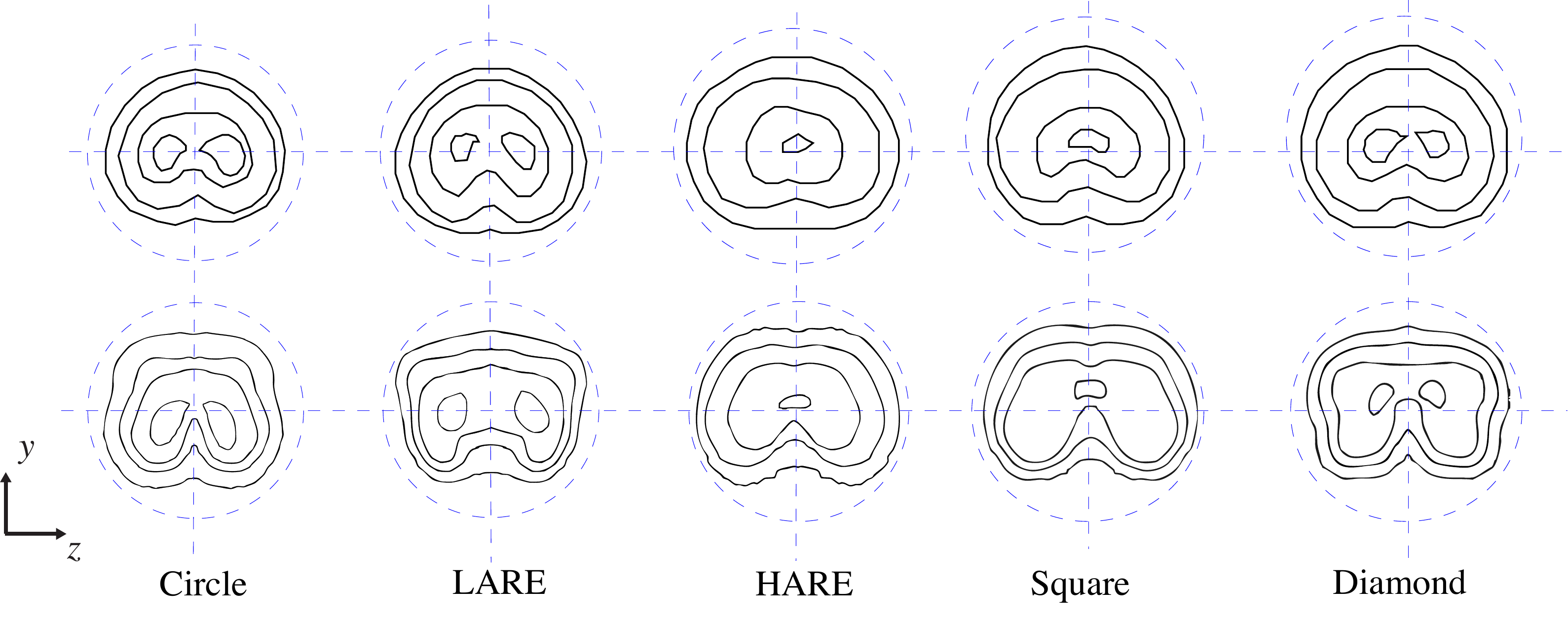}
    \caption{Contours of mixture fraction on the $y-z$ plane at $x/D = 12D$. The contours, ordered from the outermost to the innermost, correspond to $ Z=0.01,0.04,0.08,0.1$. Experiments by Salewski et al. \cite{salewski2008mixing} (top row) and LES (bottom row) for a velocity ratio $r=0.25$. }
    \label{fig:MixingPlanes}
\end{figure}

Figure~\ref{fig:MixingPlanes} shows contour lines of mixture fraction at $x/D = 12$ downstream of the nozzle exit for different nozzle geometries in the non-reacting configuration. This diagnostic enables direct comparison with the study by Salewski et al.\cite{salewski2008mixing}, who investigated mixing characteristics for circular and non-circular jets with a velocity ratio $r = 0.25$. The contours illustrate the downstream evolution of the jet core ($Z=0.1$), revealing single- or double-peak structures depending on the shape of the nozzle. Our results exhibit qualitatively similar trends, with each nozzle producing distinct core shapes ranging from double peaks in the circle, diamond, and LARE configurations to a single, centralized peak in the HARE and square nozzles. For the circle, diamond, and LARE nozzles, a pronounced counter-rotating vortex pair (CVP) induces bifurcation of the jet core, resulting in two distinct peaks in the mixture fraction contours. In contrast, the HARE and square nozzles produce a single peak, which can be attributed to enhanced interactions between the CVP and secondary vortical structures generated by the nozzle geometry, promoting additional mixing near the centerline, resulting in a more centralized scalar distribution. While Salewski et al. \cite{salewski2008mixing} utilized a flush-mounted nozzle, the present work adopts a protruding nozzle geometry (see Fig.~\ref{fig:setup}), which might contribute to small discrepancies. This geometric difference can influence the initial shear layer development and evolution of the counter-rotating vortex pair, thereby affecting the observed mixing patterns. However, overall good agreement is observed.

One of the key performance metrics used to compare these nozzles is combustion efficiency, defined as the ratio of the carbon mass conversion rate \cite{corbin2014detailed,daun2025techniques}:
\begin{equation}
\eta = \frac{\dot{m}_{\text{\rm CO}_2}}{\dot{m}_{\text{C$_\text{\rm in}$}}},
\end{equation}
where $\dot{m}_{\text{C$_\text{\rm in}$}} = \rho_j A U_j Y_{\rm C}$ is the mass flow rate of carbon entering the nozzle, held constant for a given jet velocity. Here, $A$ denotes the nozzle area and $Y_{\rm C}$ is the carbon mass fraction of the fuel. The mass flow rate of carbon dioxide exiting the computational domain, $\dot{m}_{\text{\rm CO}_2}$, is determined at five downstream exit planes (indicated in white in Fig.~\ref{fig:setup}) and averaged over time as
\begin{equation}
\dot{m}_{\text{\rm CO}_2} = \frac{1}{\tau} \int_{t_1}^{t2} \int_{A_E} \rho_{\text{\rm CO}_2} Y_{\text{\rm CO}_2} \bm{u} \cdot \hat{\bm{n}}~{\rm d}A_E\, {\rm d}t,
\end{equation}
where $\rho_{\text{\rm CO}_2}$ is the local mass density of CO$_2$, $Y_{\text{\rm CO}_2}$ is its mass fraction, $A_E$ is the total exit area, and $\bm{u} \cdot \hat{\bm{n}}$ represents the velocity component normal to the exit plane. The averaging duration, $\tau = t_2 - t_1$, is fixed at $\approx$ 10 seconds, corresponding to approximately $450D/U_j$ (with $D = 7.62~\text{cm}$ and $U_j = 3.4~\text{m/s}$), ensuring that the computed combustion efficiency is independent of temporal fluctuations. The value of $\tau$ remains constant for all flow conditions. A similar approach to compute $\eta$ was also employed by Mohit et al \cite{mohit2025impact}.

\begin{figure}[h!]
    \centering
    \includegraphics[width=3.3in]{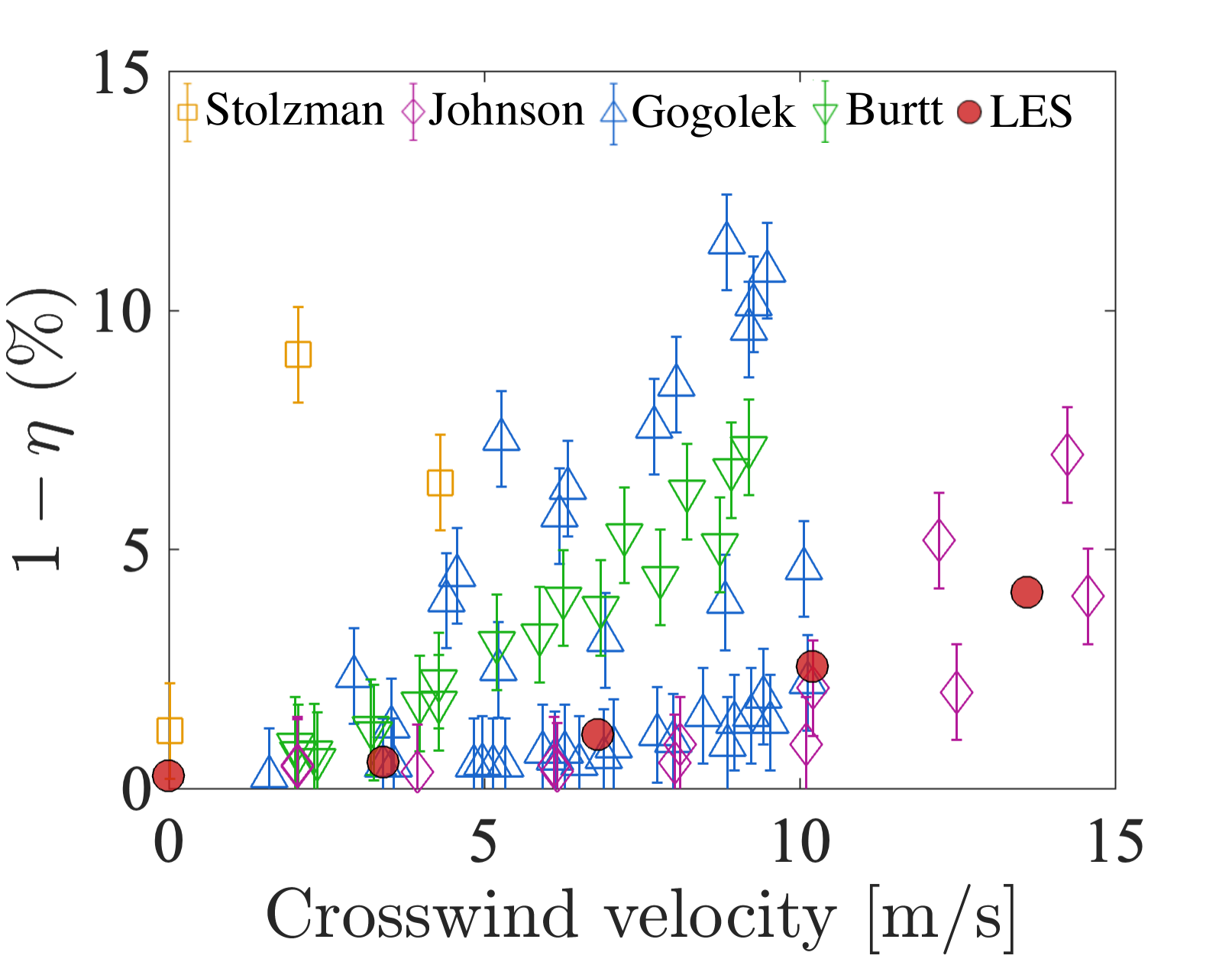}
    \caption{Comparison of combustion inefficiency as a function of crosswind velocity for the circle nozzle ($D = 7.62$ cm, $U_j = 3.4$ m/s). LES data (filled red circles). All open symbols represent flare experiment efficiency values from Stolzman et al.\cite{stolzman2025experimental} (orange squares), Gogolek\cite{gogolek2024personal} (blue triangles), Burtt\cite{burtt2014efficiency} (green inverted triangles), and Johnson and Kostuik\cite{johnson2002parametric} (magenta diamonds).}
    \label{fig:CE_compare}
\end{figure}

Experimental data, sourced from various research groups \cite{johnson2002parametric,stolzman2025experimental,burtt2014efficiency,gogolek2024personal}, provides a benchmark for evaluating the accuracy of the LES results in replicating the combustion efficiency of non-assist flare systems. 
Figure~\ref{fig:CE_compare} compares the combustion inefficiency (defined as $1-\eta$) obtained from LES  with experimental data from multiple flare studies conducted in wind tunnels and indoor calorimeters, spanning a range of crosswind velocities. 
Table~\ref{tbl:Fuels} provides the key parameters for the different experiments and the LES. The observed agreement in the overall trend of increasing inefficiency with crosswind velocity across both LES and experiments provides some validation of the simulation approach for predicting flare performance under varying crosswind conditions.

\begin{table}
  \caption{Experimental conditions of various flare studies from Gogolek\cite{gogolek2024personal}, Johnson $\&$ Kostuik\cite{johnson2002parametric}, Burtt\cite{burtt2014efficiency}, and Stolzman et al.\cite{stolzman2025experimental} using a circle nozzle. The flare efficiencies recorded from these experiments are shown in Fig.~\ref{fig:CE_compare}.}
  \label{tbl:Fuels}
  \begin{tabular}{llllllll}
    \hline
     & Johnson & Burtt & Gogolek & Stolzman & LES   \\
    \hline
    $D$ [cm]   & $2.54,~5.08$ & $7.62$ & $9.96$  & $7.62$ & $7.62$\\
    $U_j$ [m/s]   &  $2-4$ & $2$ & $0.2-22$  & $0.5$ & $3.4$\\
    $U_c$ [m/s] & $1-14.5$ & $1.5-11.5$ & $1-12.5$  & $0-5.9$ & $0-12$\\
    Fuel [$\%~\rm{CH}_4$]& $95.2\%$ & $86.03,~93.31\%$ & $95.3\%$  & $95.7\%$ & $100\%$ \\
    LHV [BTU/SCF]\textsuperscript{\emph{a}} & $907$ & $1009,~865$ & $985$  & $910$ & $920$\\
    Nozzle location [m]\textsuperscript{\emph{b}}  & $6.7$ & $5.08$ & $1.22,~2.44$ & $2.13$ & $0.99$ \\
    \hline
  \end{tabular}
  
    \textsuperscript{\emph{a}}Lower heating value of the fuel.
    \textsuperscript{\emph{b}}Position of the nozzle relative to the exit plane of the crosswind inlet.
\end{table}

Large scatter is observed in the experimental data. This variability may be due to differences in experimental setups, diagnostic techniques, and the inherent turbulence in the crosswind that is often not fully accounted for \cite{mohit2025impact}. Additionally, variations in fuel composition and dilution can further impact the measured efficiency. While the LES results fall within the experimental scatter, this highlights one of the challenges associated with making direct comparisons due to the complex and variable nature of flare operation under crosswind conditions. An extensive validation of the combustion efficiency obtained from the LES-FPVA method against similar experimental flare studies can also be found in our previous work \cite{mohit2025impact}.

%In our simulations, all boundary conditions are well defined, ensuring consistency and reproducibility across cases. However, it is important to note that real-world flare systems often operate under much less certain and more variable boundary conditions, such as fluctuations in ambient wind, fuel composition, and flow rates. Additionally, while detailed chemistry is used for the generation of the flamelet libraries, the actual chemical pathways and kinetics can be even more complex in field conditions. Although a comprehensive sensitivity analysis was beyond the scope of this study, our modeling framework and chosen conditions are validated against experimental results and provide a sound basis for evaluating the influence of nozzle geometry. We are therefore confident in the robustness of our approach for the scenarios considered here.

%%%%%%%%%%%%%%%%%%%%%%%%%%%%%%%%%%%%%%%%%%%%%%%%%%%%%%%%%%%%%%%%%%%%%%%%%%%%%%%%%%%%%%%%%%%%%%%%%%%%%%%%%%%%%%%%%%%%%%%%%%%

\section{Results}\label{sec3}
\subsection{Flame behavior}\label{subsec:5}
\begin{figure}[h!]
    \centering
    \begin{subfigure}{0.9\textwidth}
        \centering
        \includegraphics[width=0.9\textwidth]{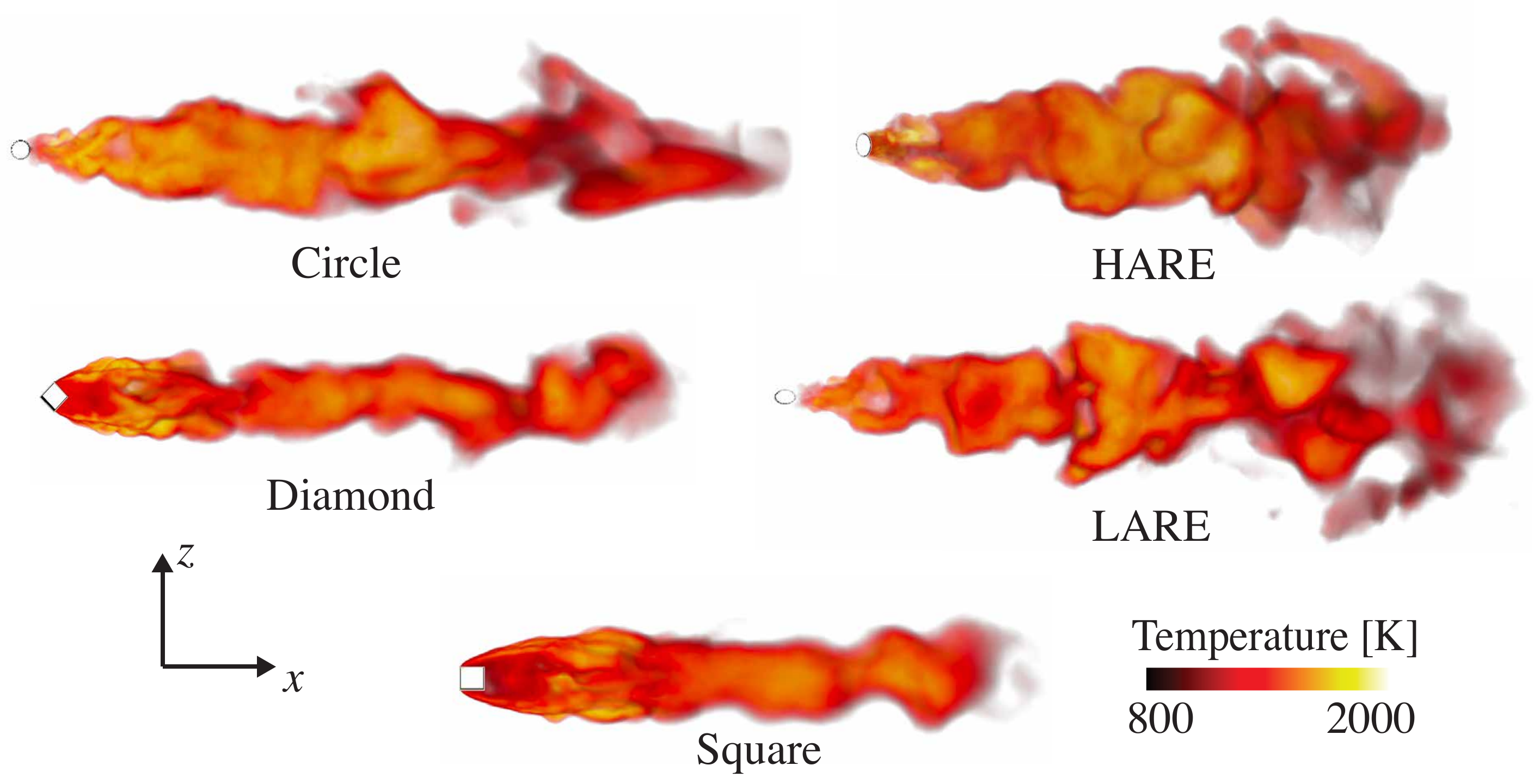}
        \caption{}
        \label{fig:flame-shapes1}
    \end{subfigure}
    \vspace{0.5em}
    \begin{subfigure}{0.85\textwidth}
        \centering
        \includegraphics[width=0.85\textwidth]{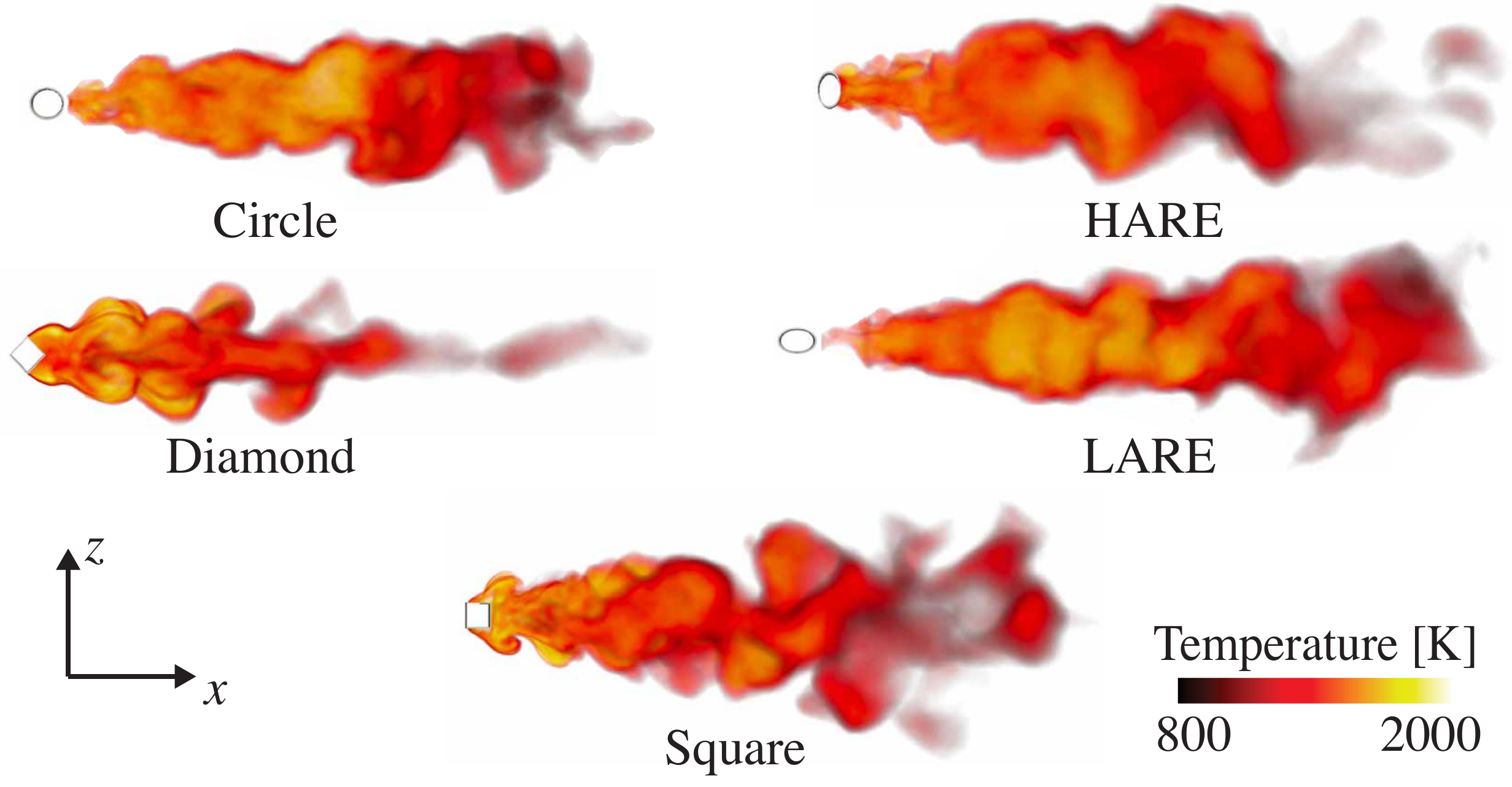}
        \caption{}
        \label{fig:flame-shapes2}
    \end{subfigure}
    \caption{Instantaneous flame profiles for various nozzle geometries (a) $U_c = 6.8$~m/s ($r=2$) and (b) $U_c = 13.6$~m/s ($r=4$), both with $U_j = 3.4$~m/s.}
    \label{fig:flame-shapes}
\end{figure}

Figure~\ref{fig:flame-shapes} displays instantaneous temperature fields for different nozzle shapes (circle, diamond, square, HARE, LARE) at two crossflow velocities. The images highlight how each geometry influences the flame’s structure, attachment, and spreading. The flame boundaries are defined using an $800$~K temperature iso-surface, following the approach of Bradley et al.\cite{bradley2016jet}. The flame length is taken as the farthest downstream location where $T=800$~K, while flame width is defined as the diameter of the cylinder enclosing the entire time-averaged $T=800$~K boundary. The internal flame structure is visualized via a volume rendering of the temperature field.

Table~\ref{tab:flame_dimensions} summarizes the average flame dimensions (time averaged for 10 seconds) and attachment zones for the different nozzles. Clear trends emerge among the nozzle geometries: the circular and LARE nozzles produce longer, narrower flames with less radial spreading, typically maintaining lengths up to $21D$ and widths as narrow as $3.5D$. These configurations exhibit more coherent jet cores, with the flame often attached on the windward side at lower crossflow ($r=2$), but tending to lift off and detach as the crosswind increases ($r=4$). In contrast, the square, diamond, and HARE nozzles yield broader, shorter flames with pronounced radial dispersion. Their flame widths are significantly larger (up to $5D$), while the overall flame length is reduced compared with rounder nozzles. In these geometries, high-temperature zones—depicted by bright red and yellow contours—are both wider and located closer to the nozzle exit. This compact, radially dispersed flame structure is a direct consequence of enhanced turbulent mixing facilitated by corner-induced separation and flow recirculation in sharp-edged geometries \cite{huser1993direct, eaton1980turbulent}.

\begin{table}
\caption{Average flame dimensions for different nozzles at $U_c = 6.8$ and $U_c = 13.6$ m/s.  Corresponding instantaneous flame profiles, average streamlines and average flame attachment are visualized in Fig.~\ref{fig:flame-shapes}-\ref{fig:flameattach}}
\begin{tabular}{lcccccc}     
    \hline
    {Nozzle shape} 
    & \multicolumn{3}{c}{$r$ = 2} 
    & \multicolumn{3}{c}{$r$ = 4} \\
    \cmidrule(lr){2-4} \cmidrule(lr){5-7} 
    & Length [m]  & Width [m] & Attachment 
    & Length [m] & Width [m] & Attachment \\
    \hline
    Circle       & 1.48 & 0.25  & Windward     &  1.14     &  0.27    &  Lifted         \\
    HARE   & 1.21  & 0.34 & Windward     &  1.01     &  0.36  & Windward      \\
    LARE   & 1.62  & 0.22 & Windward    &   1.22    &   0.23    &  Lifted          \\
    Square       & 1.18 & 0.38 & Leeward     &  0.98      & 0.40      &    Windward          \\
    Diamond      & 1.24 & 0.35 & Windward     & 1.05      & 0.38      & Windward           \\
    \hline
\end{tabular}
\label{tab:flame_dimensions}
\end{table}

\begin{figure}[h!]
    \centering
    \includegraphics[width=5in]{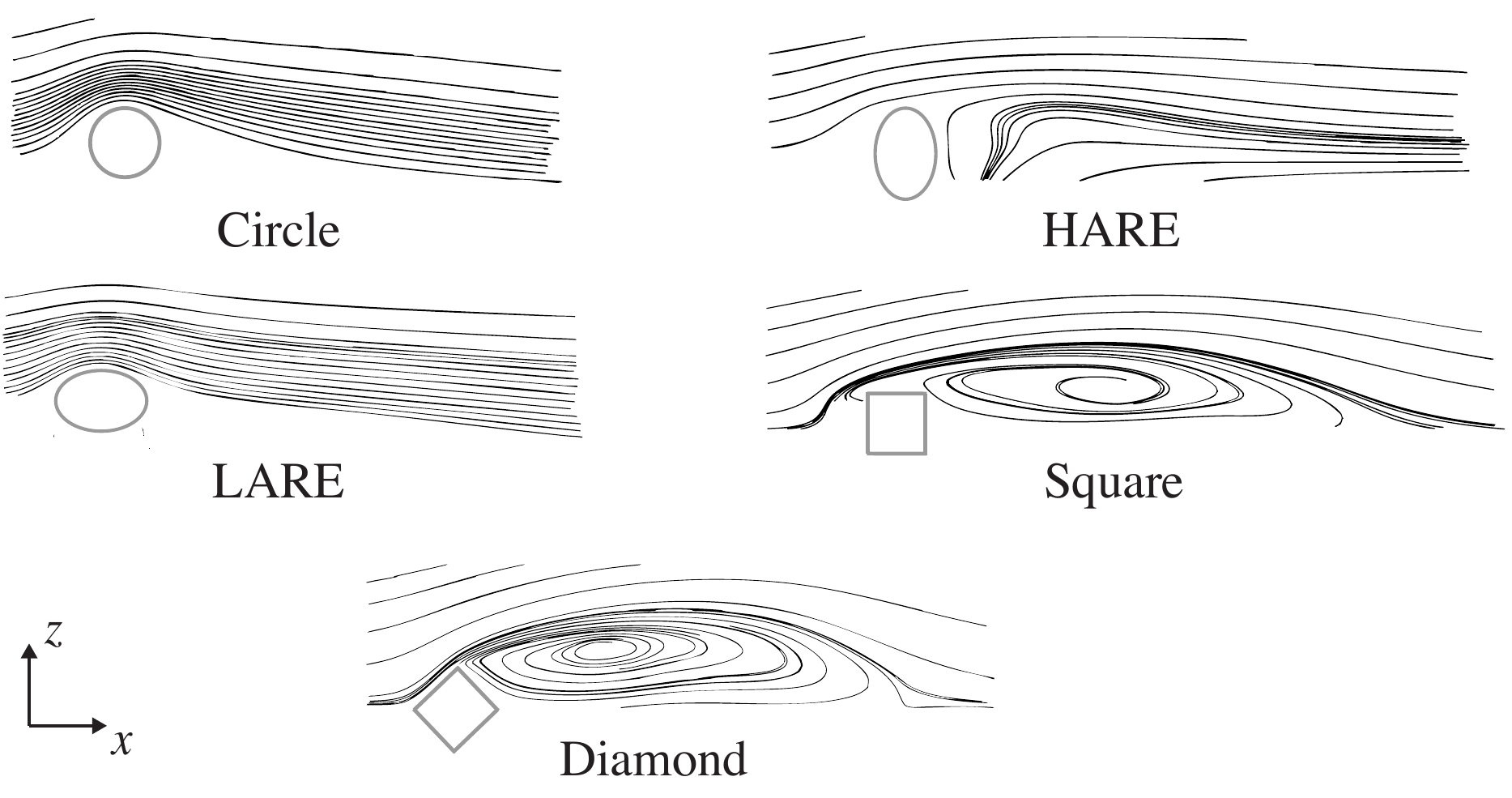}
    \caption{Streamlines of the time-averaged velocity field visualized on the $x-z$ plane at $y=3.25D$ for $U_j = 3.4$ m/s and $U_c = 6.8$ m/s.}
    \label{fig:Streamlines}
\end{figure}

Figure~\ref{fig:Streamlines} depicts the mean velocity streamlines around different nozzle geometries, highlighting the distinct flow patterns induced by each shape. The streamlines around the circle and LARE geometries remain largely attached, exhibiting minimal recirculation upstream or in the near-wake region. In contrast, the square, diamond, and HARE nozzles generate prominent recirculation zones immediately downstream of the nozzle corners, as evident by the formation of closed streamline loops and separated shear layers. These corner-induced vortices are a direct consequence of flow separation at sharp edges. Vortex shedding from corners markedly intensifies local entrainment layers and vortical structures, leading to enhanced fuel–air mixing rates. This stronger mixing advances combustion initiation, producing wider and shorter flames relative to those from smoother nozzle geometries. In contrast, circular and elliptical nozzles generate relatively axisymmetric and smooth flow profiles, which help maintain a coherent jet core and delay complete mixing, resulting in longer flames downstream. These observations are consistent with previous findings by Gutmark et al.\cite{gutmark1997passive}, who reported shorter, wider flames in rectangular nozzles due to corner-generated vortices, and by Panda et al.\cite{panda2019flame}, who demonstrated that corner-induced turbulence accelerates combustion initiation by intensifying mixing within local shear layers. 

\begin{figure}[h!]
    \centering
    \includegraphics[width=3.2in]{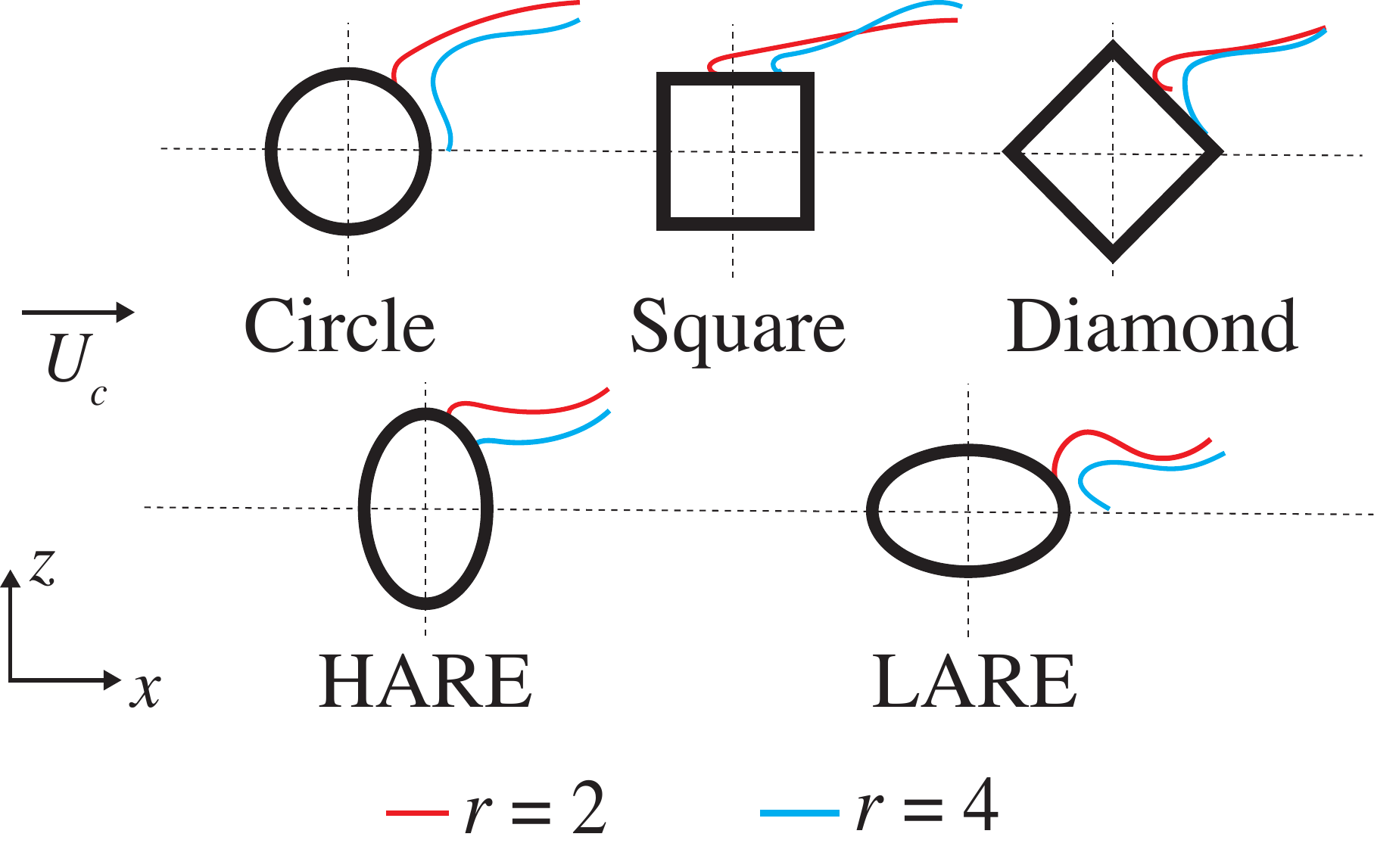}
    \caption{Comparison of the flame base for different nozzle geometries. The contours of the time-averaged temperature of $800$~K are shown for different nozzle designs. The contours are taken on the $x-z$ plane at $y = 6.5D$ (tip of the nozzle) with $U_j = 3.4$ m/s. $U_c = 6.8$ m/s (red line) and $U_c = 13.6$ m/s (blue line). Windward side (left of vertical black dashed line). Leeward side (right of vertical black dashed line).}
    \label{fig:flameattach}
\end{figure}

When examining the effects of varying velocity ratios presented in Fig.~\ref{fig:flameattach} in the form of flame attachment to the flare tip, it becomes apparent that flame stabilization and attachment dynamics differ significantly between $r=2$ and $r=4$. At $r=2$, flames are typically attached or slightly lifted, with most geometries exhibiting windward attachment. Notably, the square nozzle demonstrates minimal liftoff at this velocity ratio, maintaining attachment closer to the nozzle exit. Conversely, at higher crosswind conditions ($r=4$), the aerodynamic forces exert greater disruption on the jet, promoting shear layer instabilities and significantly influencing flame stabilization. Here, rounded nozzles (circle and LARE) experience substantial lifting, resulting in fully lifted flames due to weaker recirculation zones and lower turbulence intensity. The HARE and diamond nozzles still demonstrate partial windward attachment, while the square nozzle transitions from no liftoff at $r=2$ to partial liftoff at $r=4$. Among all configurations, the circular and LARE nozzles exhibit the largest liftoff distances under high crossflow velocity conditions ($r=4$), emphasizing their reduced flame anchoring capabilities under intensified aerodynamic disturbances. The results follow the lift-off heights trend highlighted by Haven and Kurosaka \cite{haven1997kidney}, who showed that streamlined structures tend to have higher jet-lift off than blunt structures. 

In addition to the influence of nozzle geometry and velocity ratio on flame structure and stabilization, flame attachment plays a crucial role in determining combustion efficiency. Attached flames--where the flame anchors close to the nozzle--tend to exhibit higher combustion efficiency compared with lifted or detached flames\cite{gaipl2025combustion}. This is primarily due to the presence of strong recirculation zones near the base of the flame, which facilitate the entrainment of hot combustion products back toward the nozzle and fresh reactants.

Several experimental and computational studies have reported that strong recirculation and attachment correlate with improved combustion efficiency, as they ensure rapid mixing, efficient burnout of fuel, and suppression of incomplete combustion pathways \cite{chen1998flame,schefer1998mechanism,huang2009dynamics}. Conversely, lifted flames—more commonly observed with round and smooth nozzles at higher velocity ratios—are characterized by the absence or weakening of recirculation zones. This leads to lower local temperatures near the nozzle and reduced residence times for fuel and oxidizer, allowing incomplete combustion products to escape before full oxidation can occur. The resulting impacts on combustion efficiency and the formation of incomplete combustion products are examined in detail in the following section.

%%%%%%%%%%%%%%%%%%%%%%%%%%%%%%%%%%%%%%%%%%%%%%%%%%%%%%%%%%%%%%%%%%%%%%%%%%%%%%%%%%%%%%%%%%%%%%%%%%%%%%%%%%%%%%%%%%%%%%%%%%%

\subsection{Effect of nozzle shape on combustion efficiency}\label{subsec:6}

\begin{figure}[h!]
    \centering
    \includegraphics[width=3.3in]{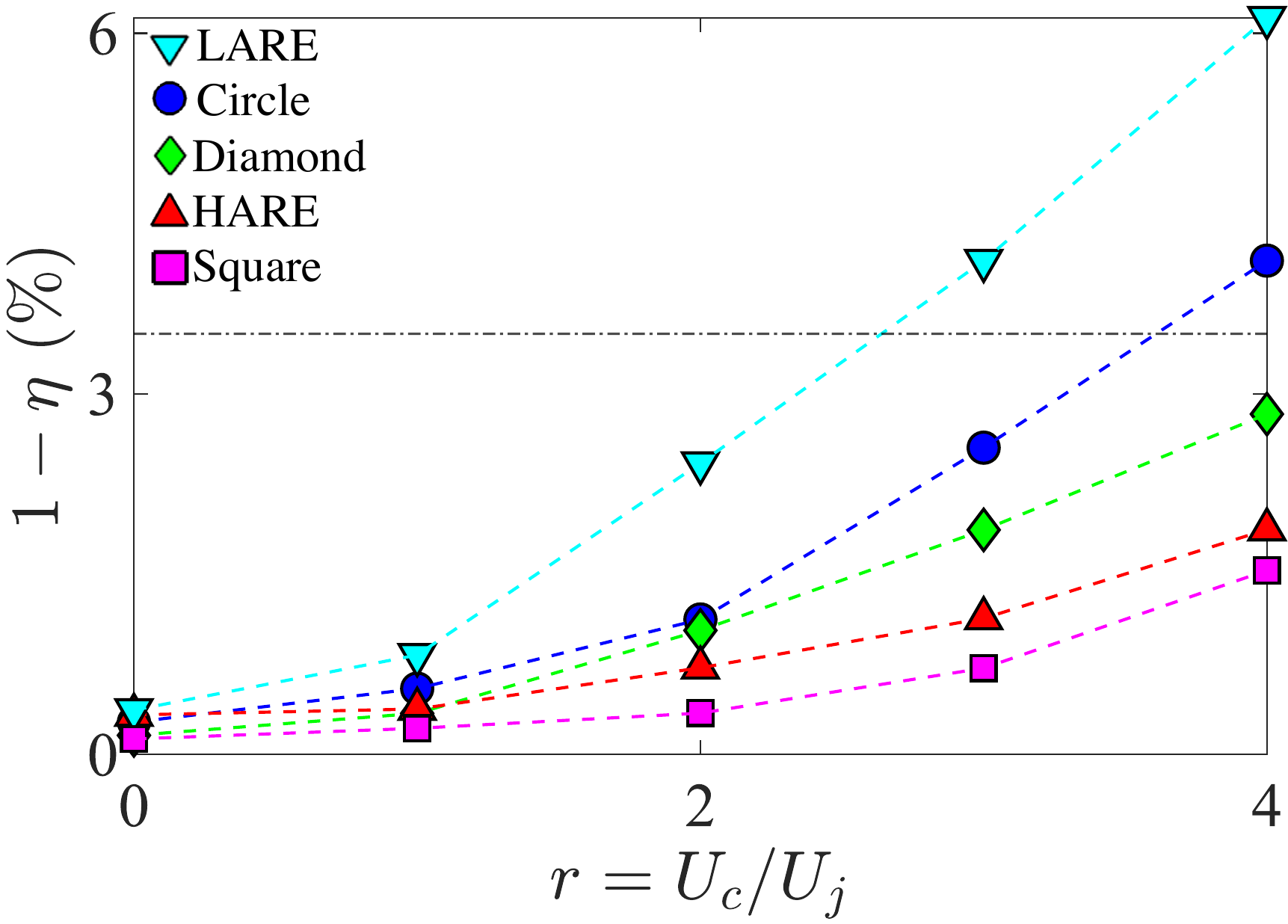}
    \caption{Combustion inefficiency as a function of velocity ratio for different nozzle shapes with fixed $U_j = 3.4$ m/s. EPA-mandated combustion efficiency ($\eta = 96.5\%$, black dotted line). LARE (cyan inverted triangle), circle (blue circle), diamond (green diamond), HARE (red triangle), square (magenta square).}
    \label{fig:ci_vs_r}
\end{figure}

Combustion efficiency serves as a key metric for assessing flare performance, reflecting the completeness of fuel conversion to CO$_2$. 
Figure~\ref{fig:ci_vs_r} shows the variation of combustion inefficiency, $1 - \eta$, with velocity ratio for different nozzle geometries at a constant jet velocity. The combustion inefficiency rises with increasing crosswind for all geometries; however, the extent of degradation depends on nozzle shape. When the crosswind and jet velocities are matched ($U_c = U_j$), geometry has only a minor effect on performance, consistent with previous observations by Stolzman et al.~\cite{stolzman2025experimental}, who reported similarly low inefficiency across multiple nozzle types under low crosswind conditions. While all nozzle shapes have similar performance at low velocity ratios (i.e., low crosswinds), the LARE and circular nozzles display the highest combustion inefficiency, particularly at the largest velocity ratio ($r=4$). Specifically, the LARE's inefficiency exceeds 6\%, compared with only 1.5\% for the square nozzle at $r=4$. This difference may seem small; however, the square nozzle would be EPA compliant ($\eta > 96.5\%$), whereas the LARE nozzle would not ($\eta < 96.5\%$). This substantial difference highlights the more pronounced sensitivity of rounded and low-aspect-ratio geometries to higher crosswind, leading to greater flame lift-off (see Fig.~\ref{fig:flameattach}) and reduced near-field recirculation as crosswind increases relative to the jet velocity. In contrast, the square nozzle consistently achieves the lowest inefficiency across all tested crosswind strengths, while the diamond and HARE geometries show intermediate performance. The critical role of nozzle-induced turbulence and recirculation in stabilizing flames and enhancing combustion efficiency is well documented in prior studies, motivating the use of strategies such as swirlers, bluff bodies, and enhanced mixing to promote robust flame anchoring and stable combustion~\cite{lefebvre2010gas, kuznetsov1990turbulence}.

The use of geometric features--such as sharp corners and non-circular nozzles--serves a fundamentally similar purpose. By deliberately shaping the nozzle to induce recirculation of hot combustion products and promote the formation of coherent vortical structures, these designs replicate the beneficial effects observed in classical swirl or bluff-body stabilized flames, but do so passively, without moving parts or added flow control devices \cite{syred1974combustion,banerjee1980flame}. 

\begin{figure}[h!]
    \centering

    % First row
    \begin{subfigure}[b]{0.31\textwidth}
        \centering
        \includegraphics[width=\textwidth]{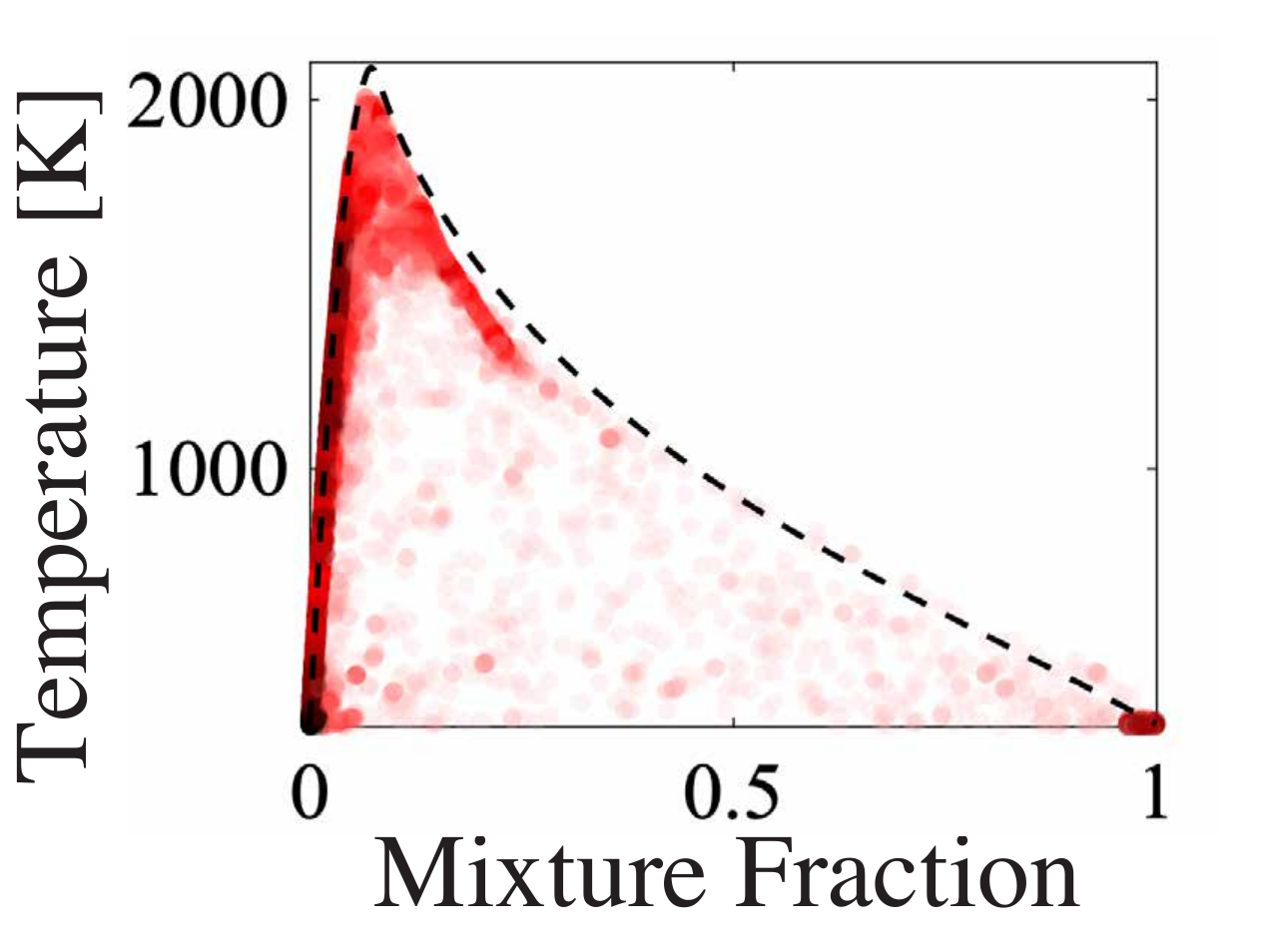}
        \caption{Square}
        \label{fig:f11sub1}
    \end{subfigure}
    \hspace{0.001\textwidth}  % Small horizontal gap
    \begin{subfigure}[b]{0.27\textwidth}
        \centering
        \includegraphics[width=\textwidth]{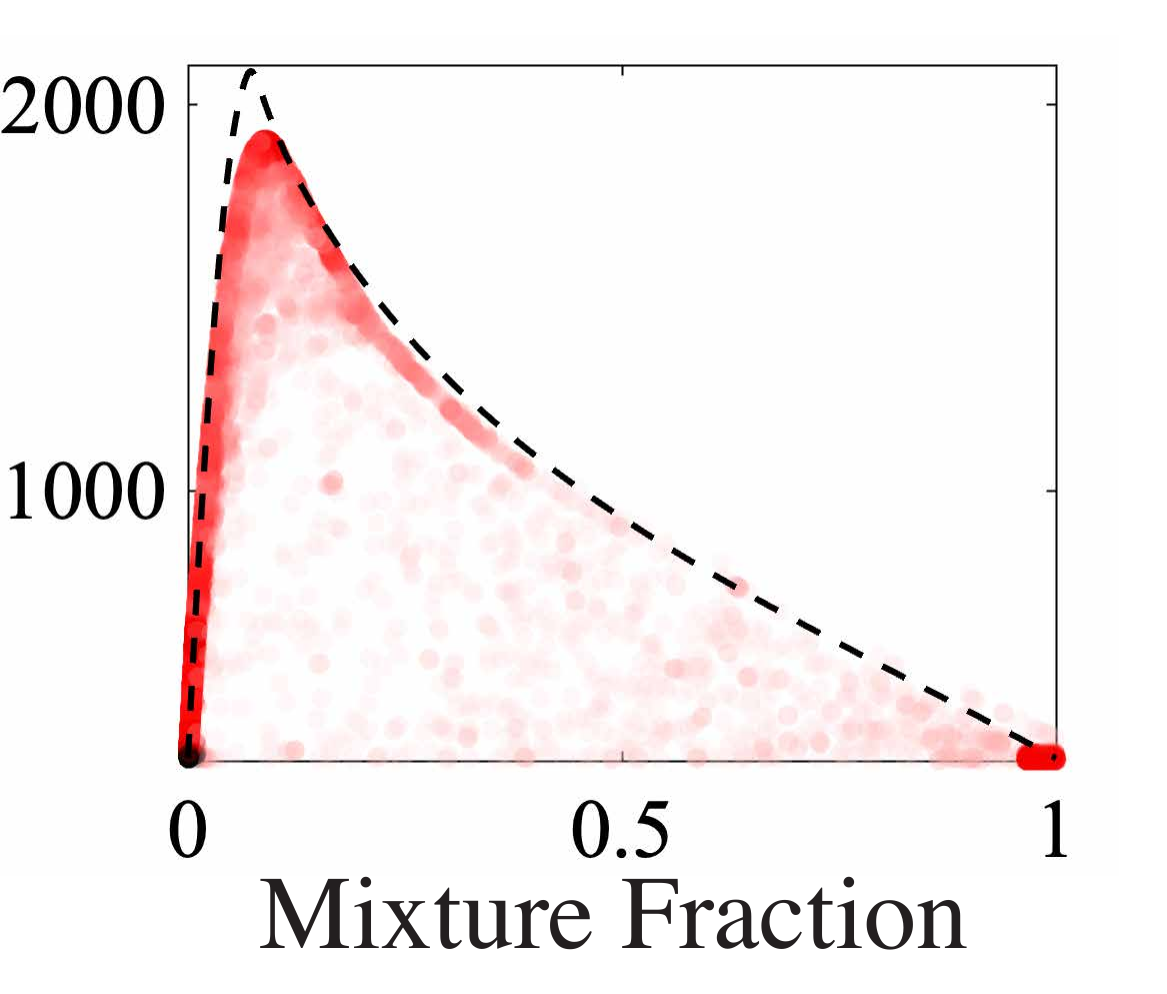}
        \caption{Circle}
        \label{fig:f11sub2}
    \end{subfigure}
    \hspace{0.001\textwidth}
    \begin{subfigure}[b]{0.35\textwidth}
        \centering
        \includegraphics[width=\textwidth]{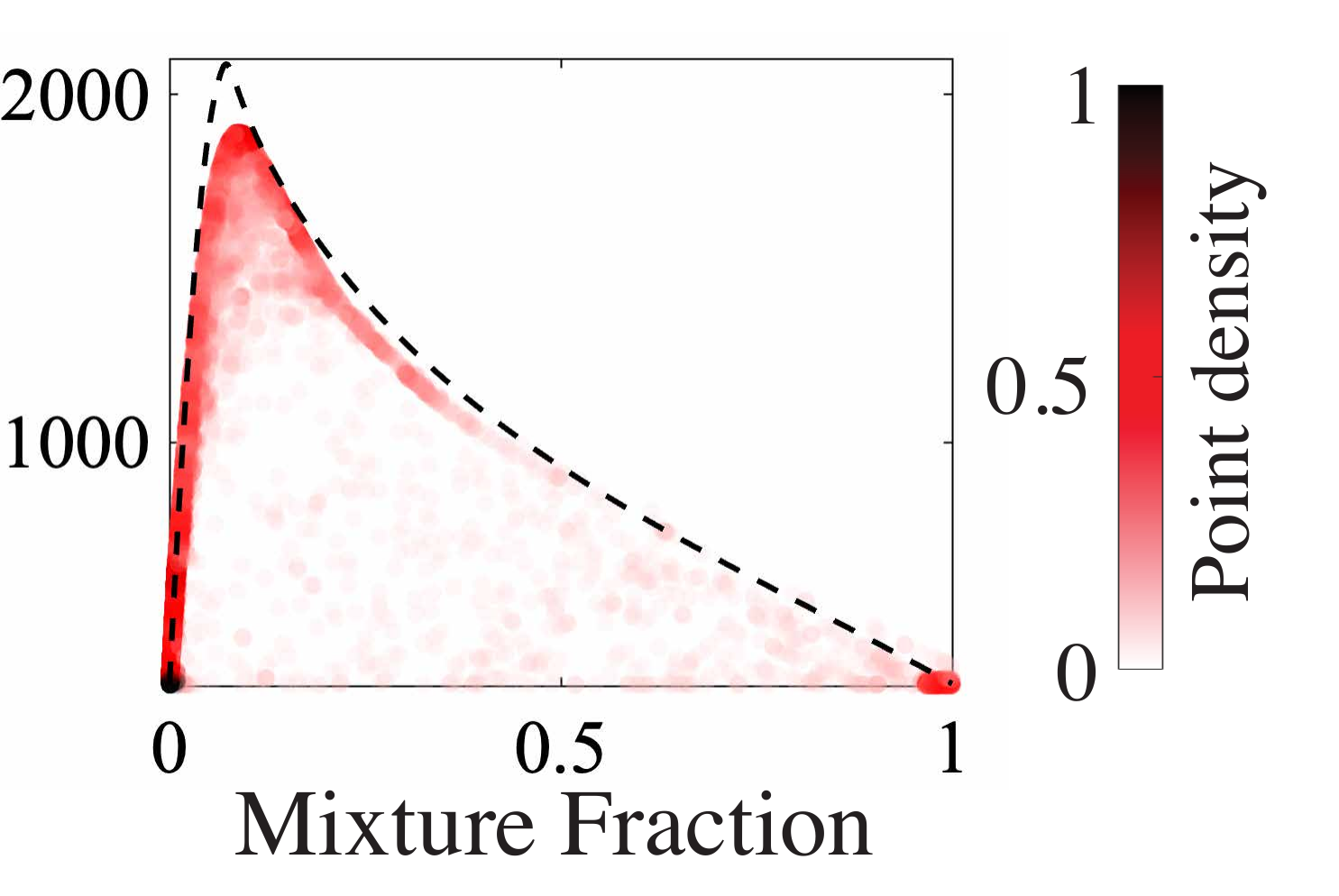}
        \caption{LARE}
        \label{fig:f11sub3}
    \end{subfigure}

    \vspace{0.5em}  % Small vertical gap

    % Second row
    \begin{subfigure}[b]{0.31\textwidth}
        \centering
        \includegraphics[width=\textwidth]{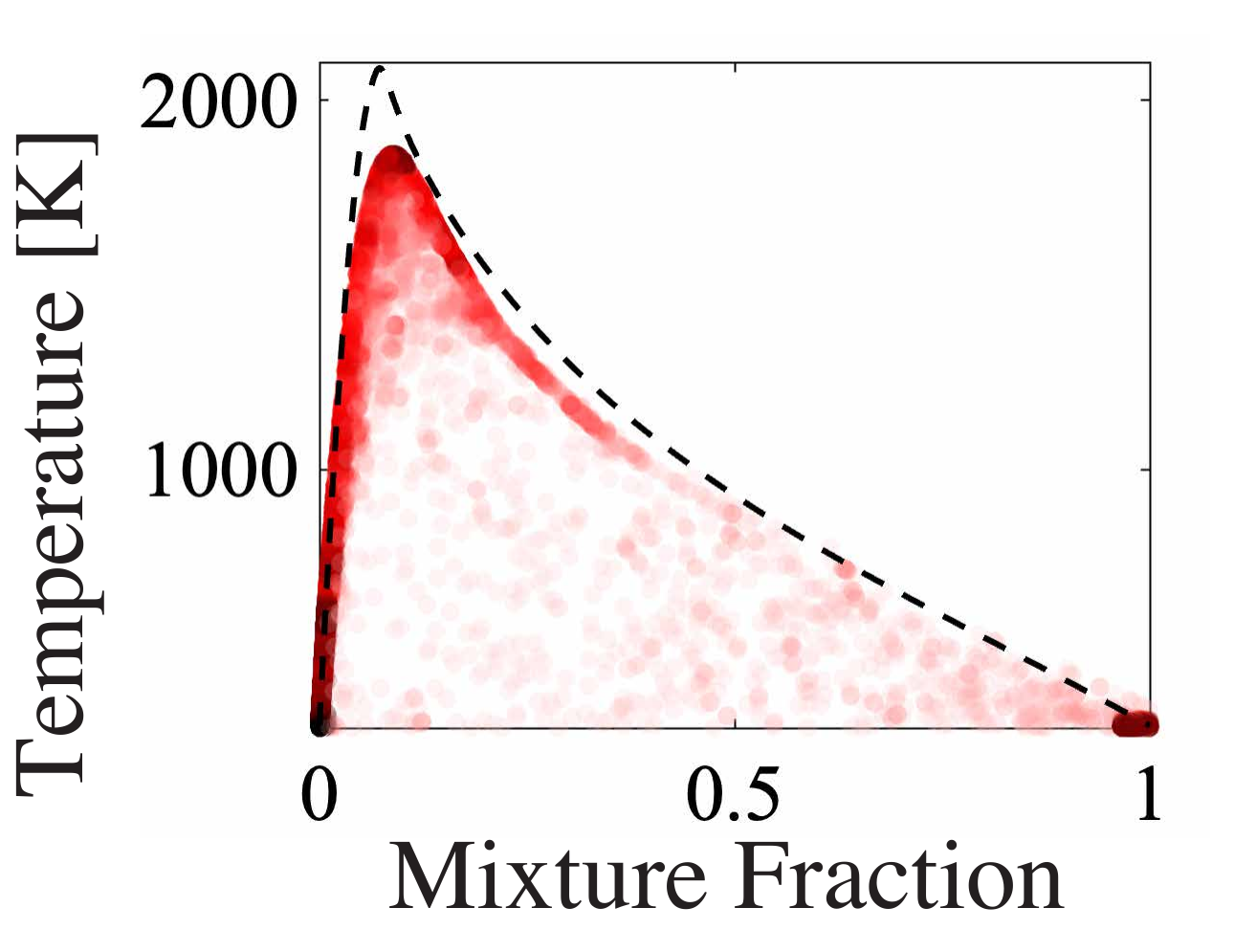}
        \caption{Square}
        \label{fig:f11sub4}
    \end{subfigure}
    \hspace{0.001\textwidth}
    \begin{subfigure}[b]{0.27\textwidth}
        \centering
        \includegraphics[width=\textwidth]{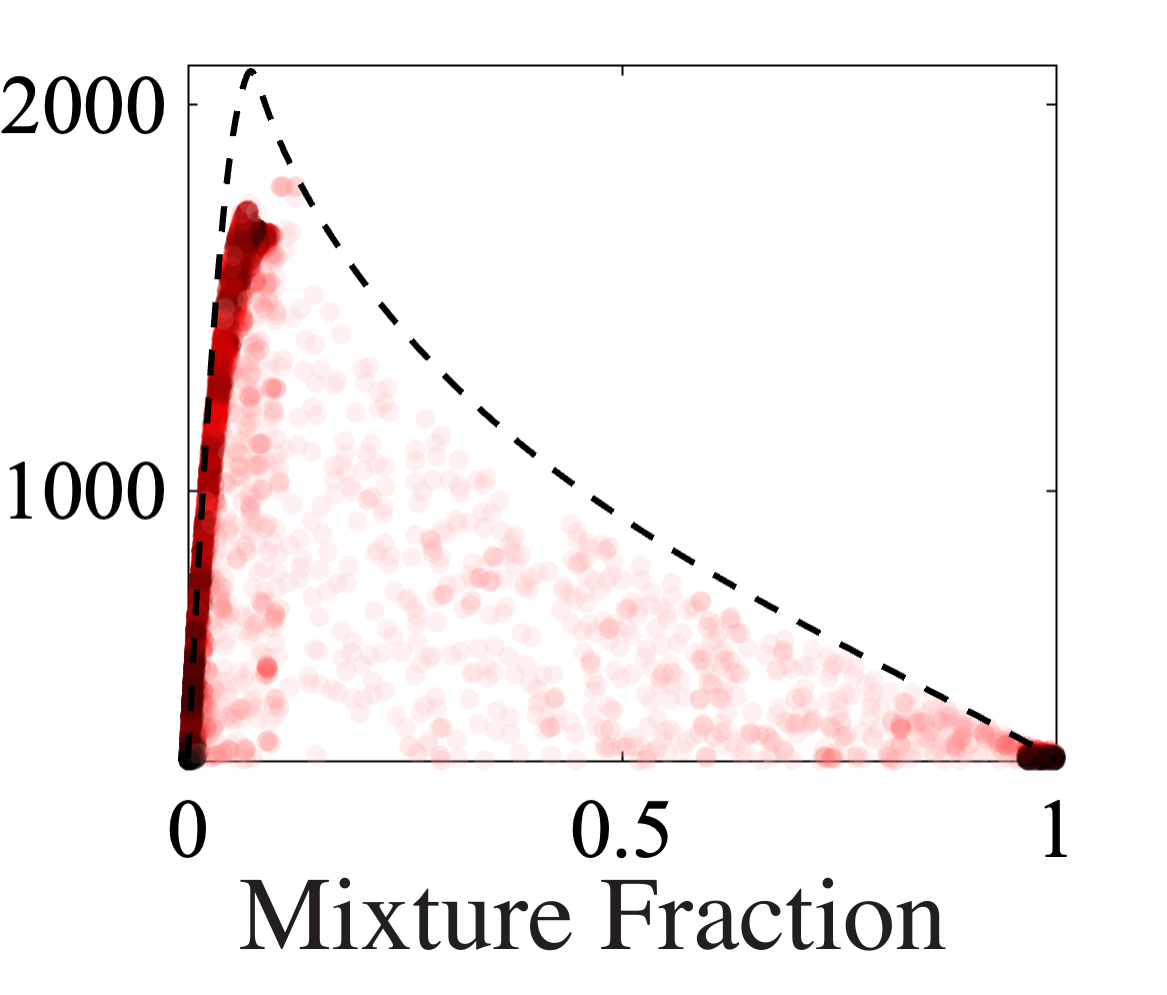}
        \caption{Circle}
        \label{fig:f11sub5}
    \end{subfigure}
    \hspace{0.001\textwidth}
    \begin{subfigure}[b]{0.35\textwidth}
        \centering
        \includegraphics[width=\textwidth]{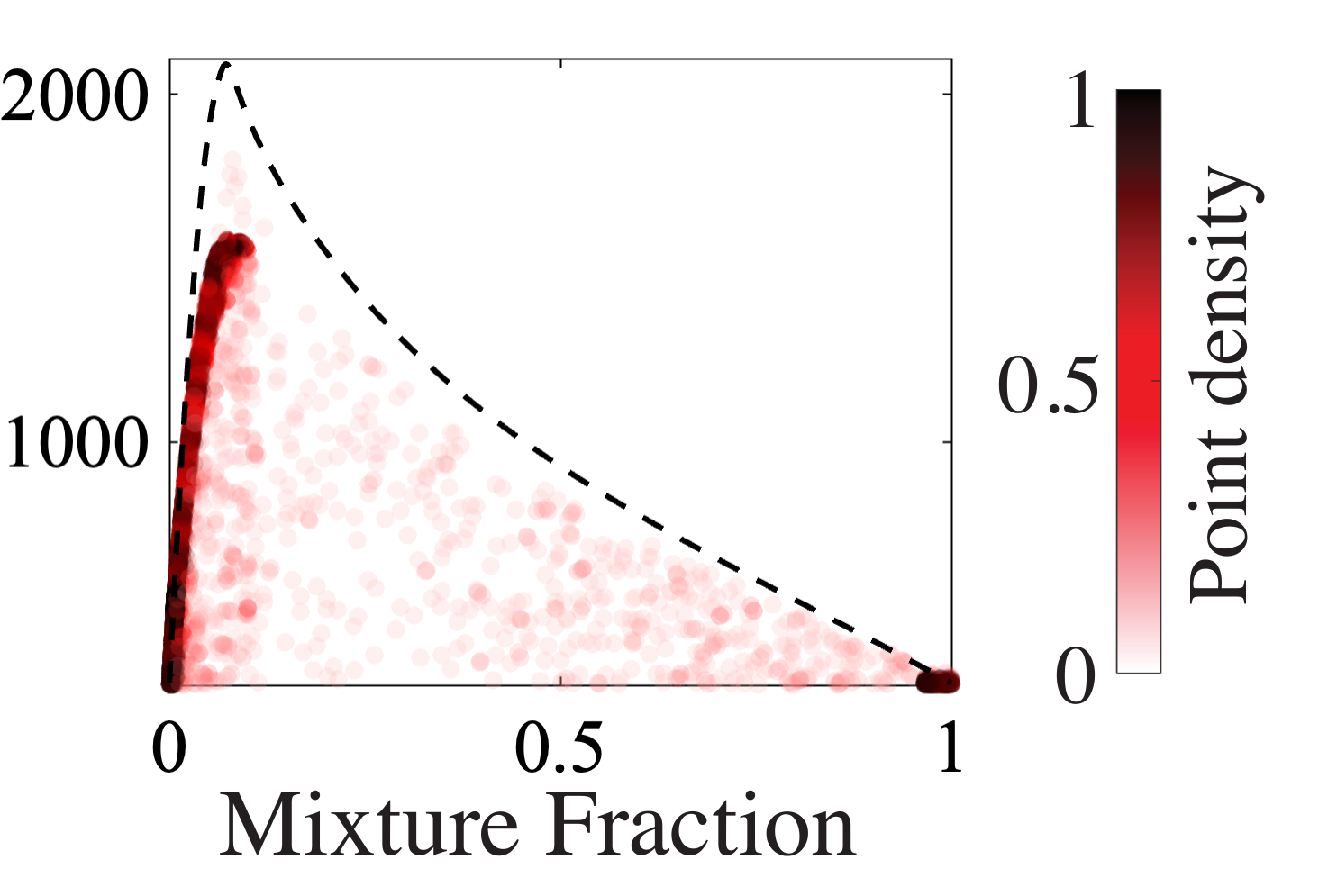}
        \caption{LARE}
        \label{fig:f11sub6}
    \end{subfigure}

    \caption{Mid-plane scatter plot of temperature as a function of mixture fraction for $U_j=3.4$ m/s, $U_c=6.8$ m/s (top row) and $U_c=13.6$ m/s (bottom row). The dashed black line is the reference solution (indicative of complete combustion) obtained from pre-tabulation of the methane reaction mechanism using \texttt{FlameMaster}.}
    \label{fig:mixfrac_temp}
\end{figure}

The observed variations in combustion inefficiency among different nozzle shapes can be explained by the scatter plots of temperature versus mixture fraction shown in Fig.~\ref{fig:mixfrac_temp}. Each subplot presents mid-plane ($z=0$) data for square, circular, and LARE nozzles at both $r=2$ and $r=4$, overlaid with the flamelet solution obtained from \texttt{FlameMaster}, where the flamelet solution is a reference for a fully chemically-equilibrated solution. The scatter points are colored based on the normalized, logarithmically--scaled population density, where a value of 1 indicates the densest clustering of points and 0 represents the sparsest regions.

A higher degree of clustering along the flamelet solution signifies that local mixing and combustion processes are more complete and closely follow equilibrium predictions. For the square nozzle, points are densely clustered near the flamelet solution for both velocity ratios, particularly at $r=2$, indicating efficient fuel-oxidizer mixing and reaction completeness. This is consistent with the high combustion efficiency observed for the square nozzle in crosswind, as efficient recirculation and anchoring help maintain high flame temperatures and minimize deviations from complete combustion.

In contrast, the circular and LARE nozzles exhibit broader scatter, with more points deviating below the fully chemically-equilibrated solution—particularly at higher velocity ratios ($r=4$). These deviations reflect greater spatial and temporal inhomogeneities, reduced peak temperatures, and increased prevalence of partially reacted or unmixed pockets.  Such behavior is indicative of incomplete combustion, as portions of the fuel-air mixture fail to be completely oxidized before being transported downstream. Unreacted pockets of fuel prevalent in downstream regions have been shown to be a major source of inefficiency in the recent study by Gaipl et al.~\cite{gaipl2025combustion}. 
The spatial distribution of temperature-mixture fraction data offers a mechanistic explanation for the observed efficiency trends: nozzles that promote tighter clustering around the flamelet solution (such as square) achieve higher combustion efficiency, while those with broader scatter (such as LARE) experience elevated inefficiency due to incomplete mixing and reaction. 

\begin{figure}[h!]
    \centering
    % Top row: orientation angle schematic centered
    \begin{subfigure}[b]{0.5\textwidth}
        \centering
        \includegraphics[width=\textwidth]{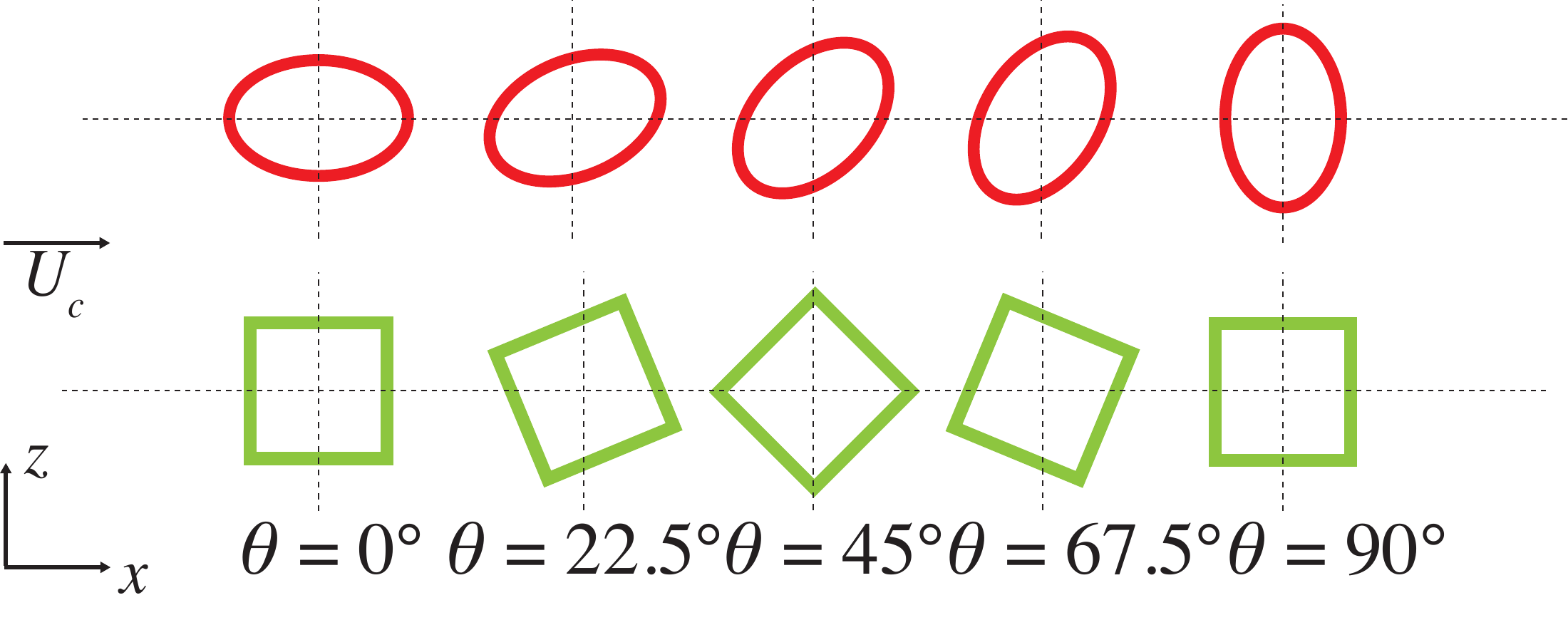}
        %\caption{}
        \label{fig:orientation}
    \end{subfigure}
    
    % Small vertical gap
    \vspace{0.8em}
    
    % Bottom row: two line plots side by side
    \begin{subfigure}[b]{0.36\textwidth}
        \centering
        \includegraphics[width=\textwidth]{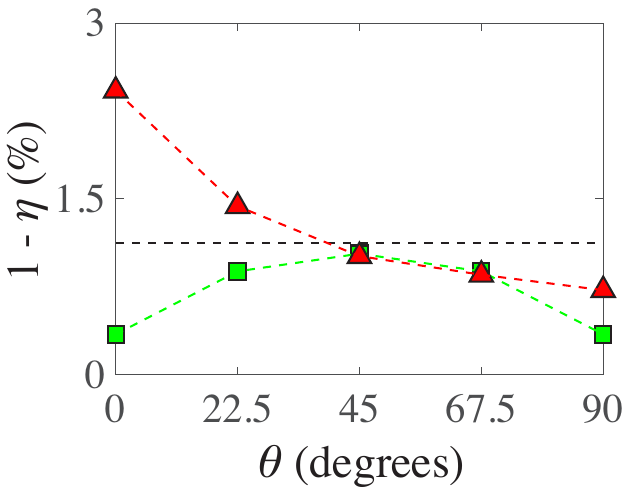}
        \caption{$r=2$}
        \label{fig:2r}
    \end{subfigure}%
    \hspace{1em}
    \begin{subfigure}[b]{0.325\textwidth}
        \centering
        \includegraphics[width=\textwidth]{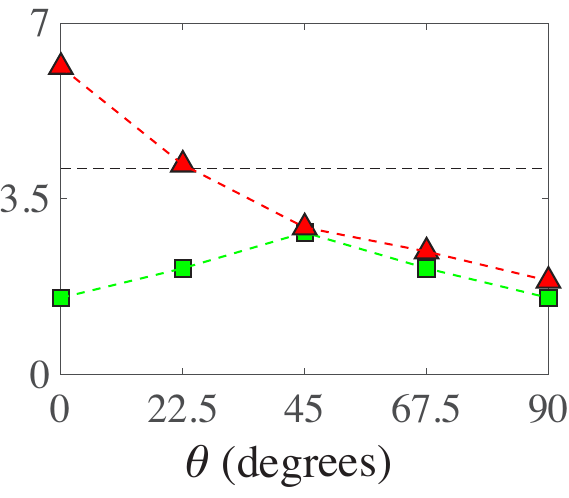}
        \caption{$r=4$}
        \label{fig:4r}
    \end{subfigure}
    \caption{Combustion inefficiency as a function of crosswind angle ($\theta$) for different nozzle shapes with $U_j = 3.4$ m/s for $U_c = 6.8$ m/s and $U_c = 13.6$ m/s. Circle nozzle (black dotted line), square nozzle (green square), and elliptic nozzle (red triangle).}
    \label{fig:angle_sweep}
\end{figure}

So far, each nozzle shape was subjected to uni-directional crosswind. In real settings, speed and turbulent intensity are uncontrollable and variable. Additional tests were conducted to assess the sensitivity of combustion inefficiency to crosswind orientation for the non-symmetric nozzles, i.e., both elliptical (red triangles) and square (green squares) nozzles. The different orientations and combustion inefficiency results are presented in Fig.~\ref{fig:angle_sweep}. The combustion inefficiency for the circular nozzle (which is invariant with crosswind direction) is provided as a benchmark for comparison. The results indicate that, across all tested angles ($\theta = 0^\circ$ to $90^\circ$), the square nozzle consistently achieves lower or comparable combustion inefficiency relative to the circular reference, highlighting its robust performance irrespective of wind orientation. In contrast, the elliptical nozzle generally performs worse than the circular reference at lower angles but approaches or even surpasses the circular nozzle at higher orientations ($\theta \geq 67.5^\circ$, from LARE to HARE). Overall, these findings highlight that the square nozzle design delivers the highest combustion efficiency across all crosswind directions.

%%%%%%%%%%%%%%%%%%%%%%%%%%%%%%%%%%%%%%%%%%%%%%%%%%%%%%%%%%%%%%%%%%%%%%%%%%%%%%%%%%%%%%%%%%%%%%%%%%%%%%%%%%%%%%%%%%%%%%

\subsection{Effect of nozzle shape on mixing and strain}\label{subsec:7}

\begin{figure}[h!]
    \centering
    \includegraphics[width=3.3in]{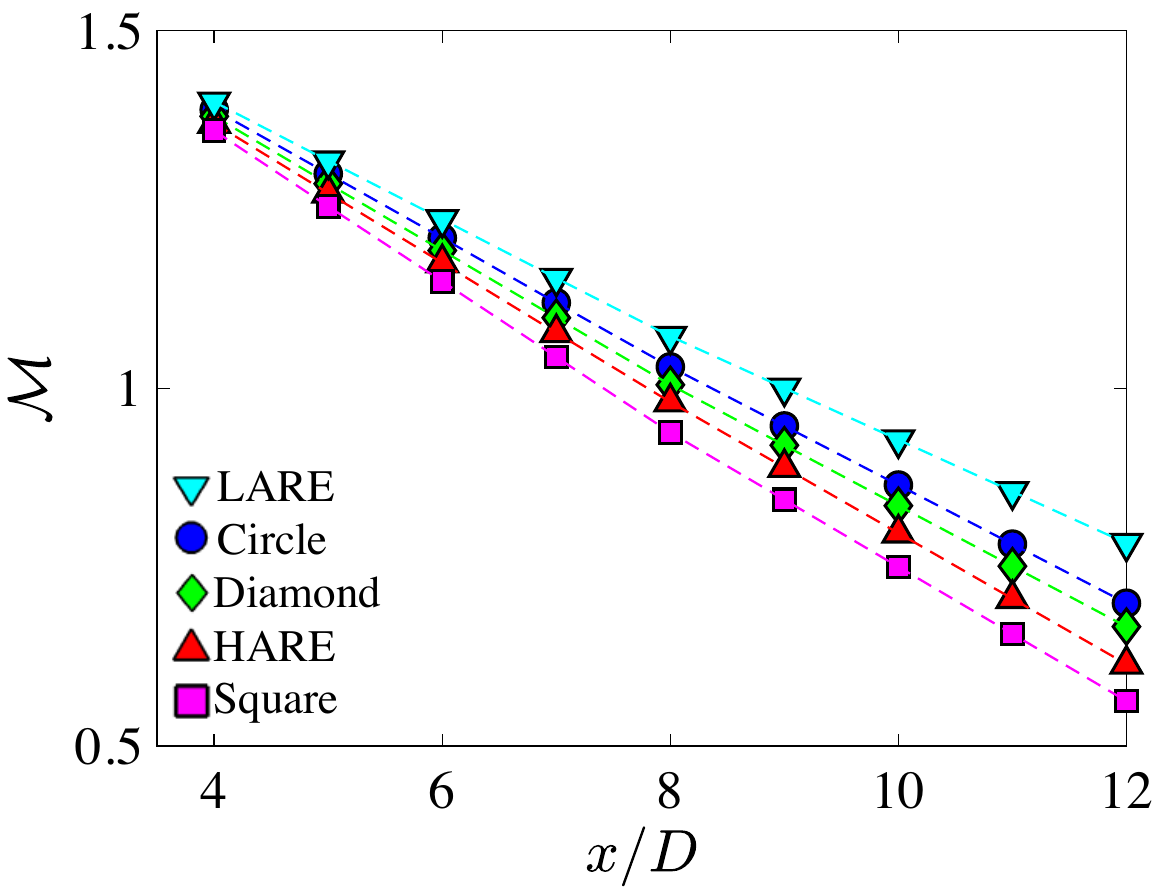}
    \caption{Composition mixing parameter $\mathcal{M}$ as a function of axial distance for various nozzle shapes. LARE (cyan inverted triangle), circle (blue circle), diamond (green diamond), HARE (red triangle), square (magenta square).}
    \label{fig:mixingparam}
\end{figure}

Figure~\ref{fig:MixingPlanes} offers a qualitative perspective on local mixing characteristics for the various nozzle designs in non-reacting flow; however, it does not fully capture the overall, global mixing behavior. For this, we employ a mixing parameter  $\mathcal{M}$, given by \cite{salewski2008mixing}
\begin{equation}
    \mathcal{M}(x)=\frac{\sqrt{Z_{\rm{var}}}}{{\langle Z\rangle}},
\end{equation}
where $Z_{\rm{var}}$ is the spatial variance of mixture fraction, given by 
\begin{equation}
    Z_{\rm{var}}(x)=\frac{1}{A_z}\int(Z-{\langle Z\rangle})^2dA_z,
\end{equation}
where $A_z$ encloses all areas of $Z$ $\geq$ 0.01 in a given plane and $\langle~\rangle$ represents mean quantities.
$\mathcal{M} = 0$ indicates a perfectly compositionally homogeneous mixture. Figure \ref{fig:mixingparam} shows $\mathcal{M}$ computed in the $y-z$ plane as a function of the axial distance. The gradient of each curve reflects the rate at which mixing occurs: a steeper decline corresponds to faster homogenization. Across all cases, $\mathcal{M}$ decreases monotonically with downstream distance; however, the rate and extent of mixing vary appreciably between nozzle designs. 

Among the tested configurations, the LARE nozzle consistently mixes most slowly (highest $\mathcal{M}$ and shallowest slope), while the square nozzle exhibits the most rapid mixing (lowest $\mathcal{M}$ and steepest slope). Circle, diamond, and HARE nozzles fall in between. These trends underscore the strong effect of nozzle shape on mixing: geometries with pronounced corners (such as the square) enhance turbulent fluctuations and shear, promoting rapid homogenization through increased vortex shedding and secondary structures~\cite{liscinsky1996crossflow,salewski2008mixing}. In contrast, smoother profiles (LARE, circle) maintain a more coherent jet core, delaying complete mixing.

\begin{figure}[h!]
    \centering
    \includegraphics[width=4.4in]{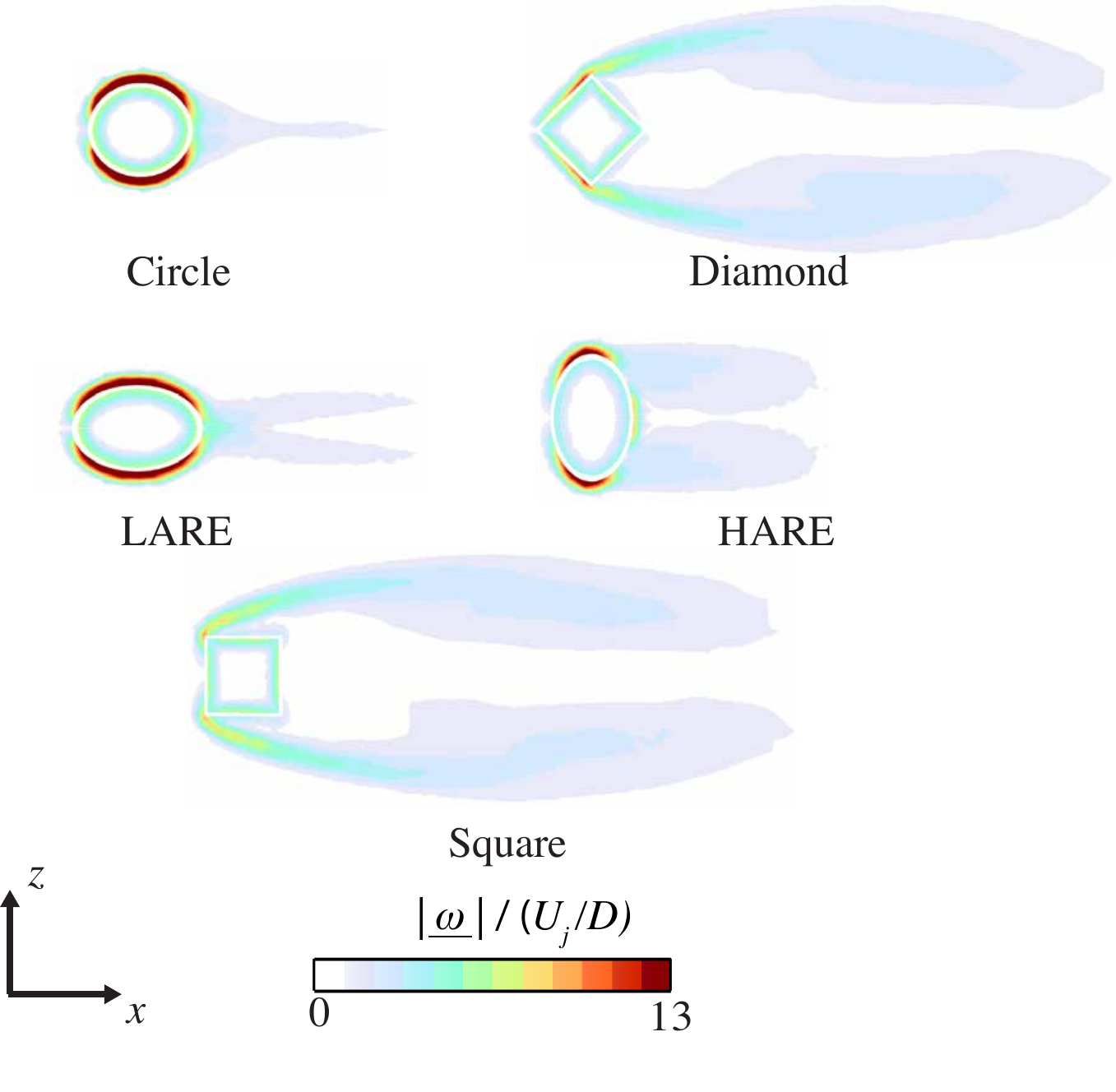}
    \caption{Comparison of vorticity magnitude for various nozzle geometries. The values are shown on the $x-z$ plane at $y = 3.25D$ with $U_j = 3.4$ m/s and $U_c = 13.6$ m/s. }
    \label{fig:vorticity}
\end{figure}

Figure~\ref{fig:vorticity} presents contours of vorticity magnitude taken along the $x-z$ plane for all nozzle geometries. Distinct variations in vorticity structure and magnitude are evident among the different designs. For the square and diamond nozzles, intense vorticity is concentrated along the sharp corners, rapidly spreading into the downstream flow and promoting the breakdown of the jet core. This enhanced generation of shear layers and vortical structures fosters vigorous turbulent mixing and facilitates the rapid homogenization of the scalar field, consistent with the lower values of the mixing parameter $\mathcal{M}$ shown in Fig.~\ref{fig:mixingparam}.

In contrast, the circular and LARE nozzles display relatively smooth and symmetric vorticity distributions confined near the jet boundary, with substantially weaker extension into the downstream region. The reduced intensity and spatial reach of vorticity in these cases correspond to a more stable jet core and delayed transition to turbulence, leading to slower mixing and greater persistence of unmixed regions downstream. The HARE nozzle exhibits an intermediate behavior, with some enhancement of vorticity relative to the circle and LARE designs but less pronounced than in the square or diamond cases.

\begin{figure}[h!]
    \centering

    % First row
    \begin{subfigure}[b]{0.29\textwidth}
        \centering
        \includegraphics[width=\textwidth]{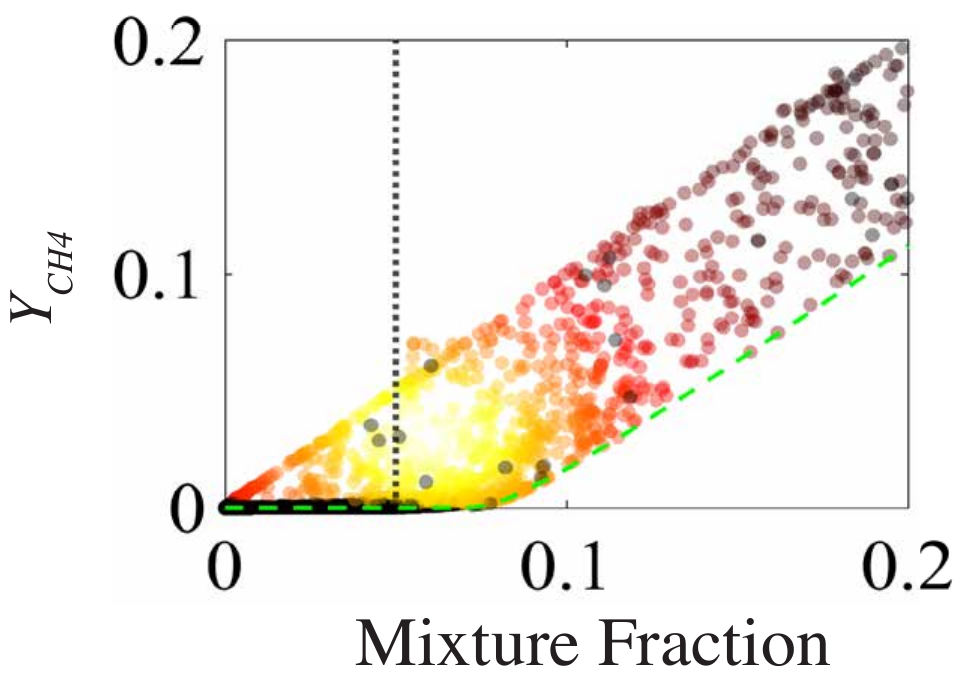}
        \caption{Circle}
        \label{fig:sub1}
    \end{subfigure}
    \hspace{0.001\textwidth}  % Small horizontal gap
    \begin{subfigure}[b]{0.253\textwidth}
        \centering
        \includegraphics[width=\textwidth]{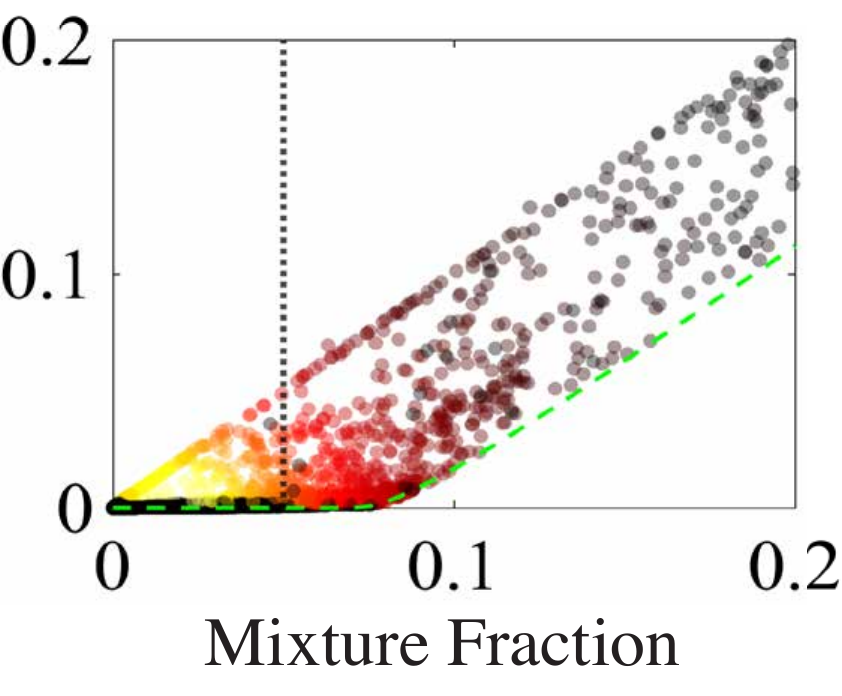}
        \caption{Square}
        \label{fig:sub2}
    \end{subfigure}
    \hspace{0.001\textwidth}
    \begin{subfigure}[b]{0.33\textwidth}
        \centering
        \includegraphics[width=\textwidth]{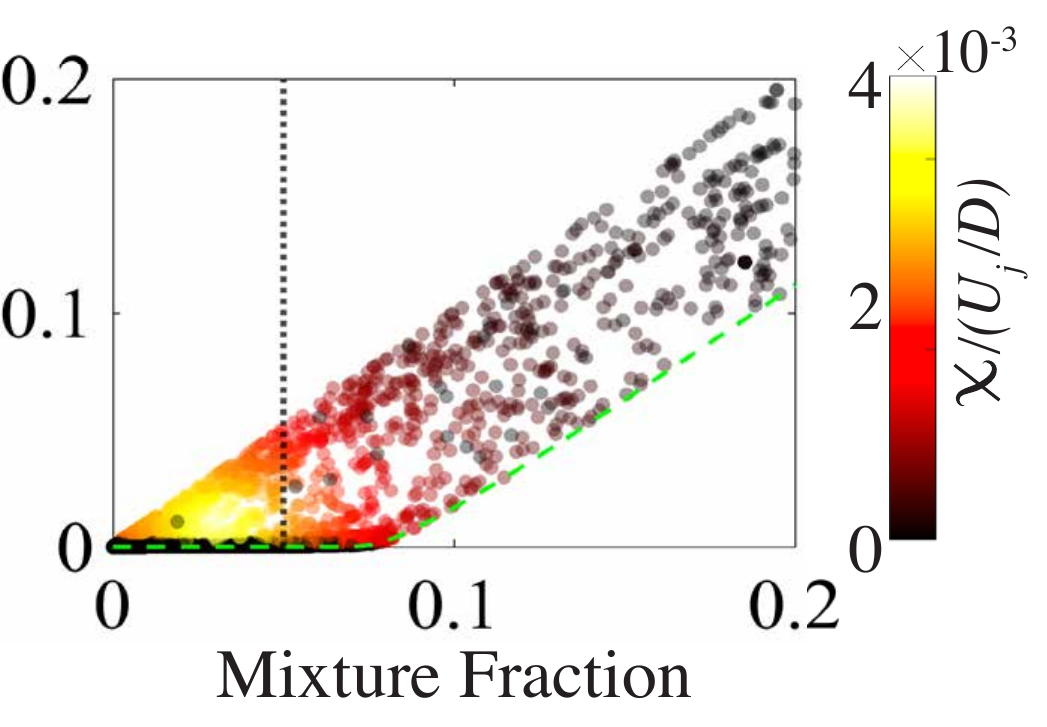}
        \caption{Diamond}
        \label{fig:sub3}
    \end{subfigure}

    \vspace{0.5em}  % Small vertical gap

    % Second row
    \begin{subfigure}[b]{0.29\textwidth}
        \centering
        \includegraphics[width=\textwidth]{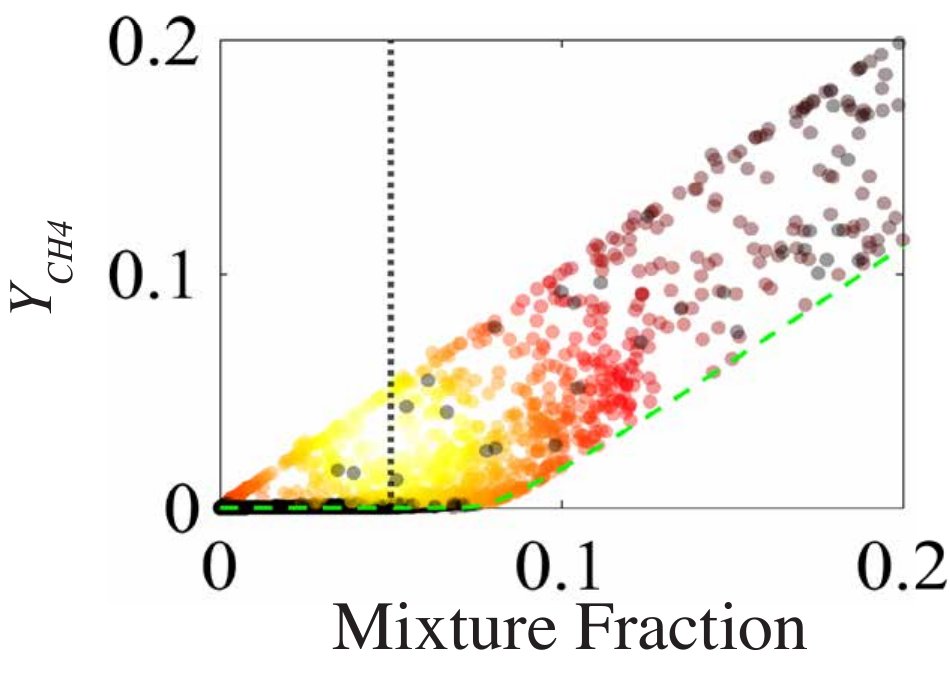}
        \caption{LARE}
        \label{fig:sub4}
    \end{subfigure}
    \hspace{0.001\textwidth}
    \begin{subfigure}[b]{0.32\textwidth}
        \centering
        \includegraphics[width=\textwidth]{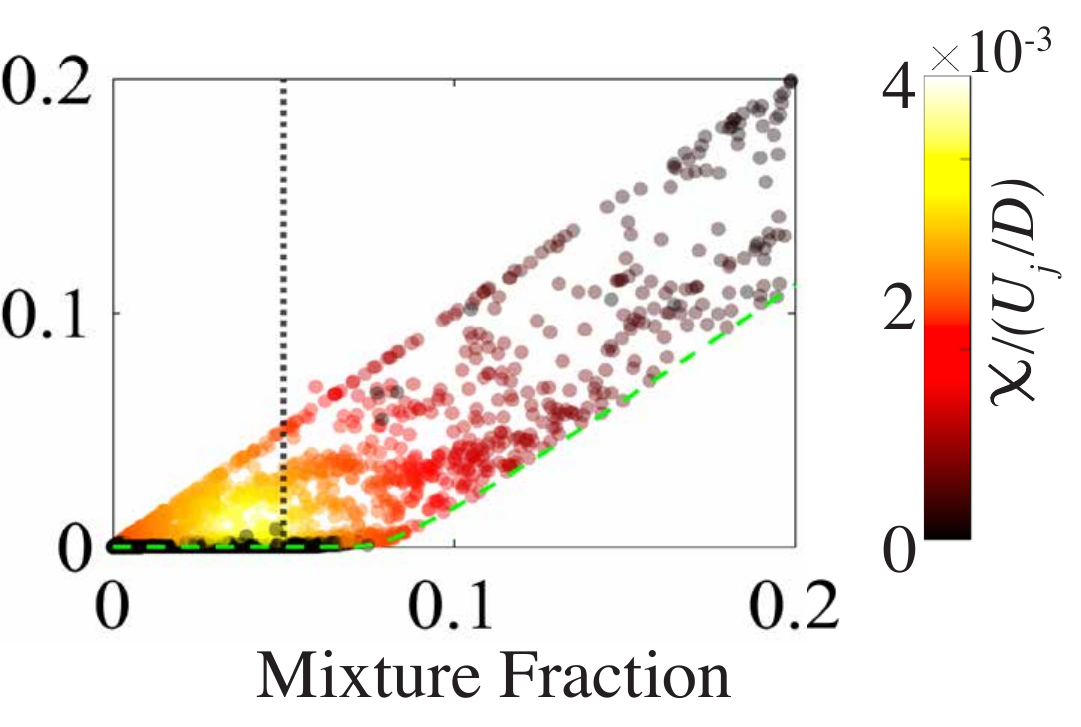}
        \caption{HARE}
        \label{fig:sub5}
    \end{subfigure}

    \caption{Mid-plane scatter plot of methane mass fraction as a function of mixture fraction for various nozzle geometries colored by the normalized mean scalar dissipation rate (normalized by $U_j/D$) for $U_j=3.4$ m/s and $U_c=13.6$ m/s. The dashed green-line is the reference solution obtained during pre-tabulation from \texttt{FlameMaster}. The dotted black-line is the stoichiometric mixture fraction for methane.}
    \label{fig:scatter_dissipation}
\end{figure}

Figure~\ref{fig:scatter_dissipation} shows mid-plane scatter plots of methane mass fraction versus mixture fraction for each nozzle geometry, colored by the local normalized mean scalar dissipation rate. The dashed green line indicates the equilibrium solution from \texttt{FlameMaster}, while the dotted black line marks the stoichiometric mixture fraction.

In all cases, points are bounded between the upper limit ($Y_{\mathrm{CH}_4} = Z$) and the lower equilibrium limit from \texttt{FlameMaster}. Deviation below the upper limit towards the green curve reflects methane consumption due to reactions. Scalar dissipation rates peak near the stoichiometric region, highlighting intense local mixing and flame straining across all nozzle geometries under high crosswind.

Despite this commonality, the spatial distribution and clustering of points reveal important differences in the response to strain. For square, diamond and HARE nozzles, the majority of points remain tightly clustered near or just before the stoichiometric line. This suggests that enhanced recirculation and localized mixing, driven by corner-induced turbulence, help alleviate the adverse effects of elevated scalar dissipation. The presence of robust recirculation zones in these geometries acts to entrain hot products and stabilize the reaction zone, mitigating the potential for incomplete combustion despite the high imposed strain. Notably, the peaks of maximum scalar dissipation remain closely associated with these high-density clusters, indicating efficient and resilient mixing behavior.

Conversely, the circle and LARE nozzles exhibit broader, more dispersed distributions of methane mass fraction, with elevated scalar dissipation rates persisting over a wider range of mixture fractions. The absence of strong recirculation in these smoother, more axisymmetric geometries reduces their ability to counteract strain-induced flame thinning. As a result, these configurations are more susceptible to flame stretching, incomplete combustion, and efficiency losses (see Fig.~\ref{fig:ci_vs_r}),  greater scatter and deviation from the reference solution. While high crosswind strain affects all nozzles, the ability to sustain recirculation and rapid mixing plays a critical role in maintaining high combustion efficiency and limiting fuel slip.

%%%%%%%%%%%%%%%%%%%%%%%%%%%%%%%%%%%%%%%%%%%%%%%%%%%%%%%%%%%%%%%%%%%%%%%%%%%%%
\subsection{Flame blow-out velocity}\label{subsec:8}

\begin{figure}[h!]
    \centering
    \includegraphics[width=3.3in]{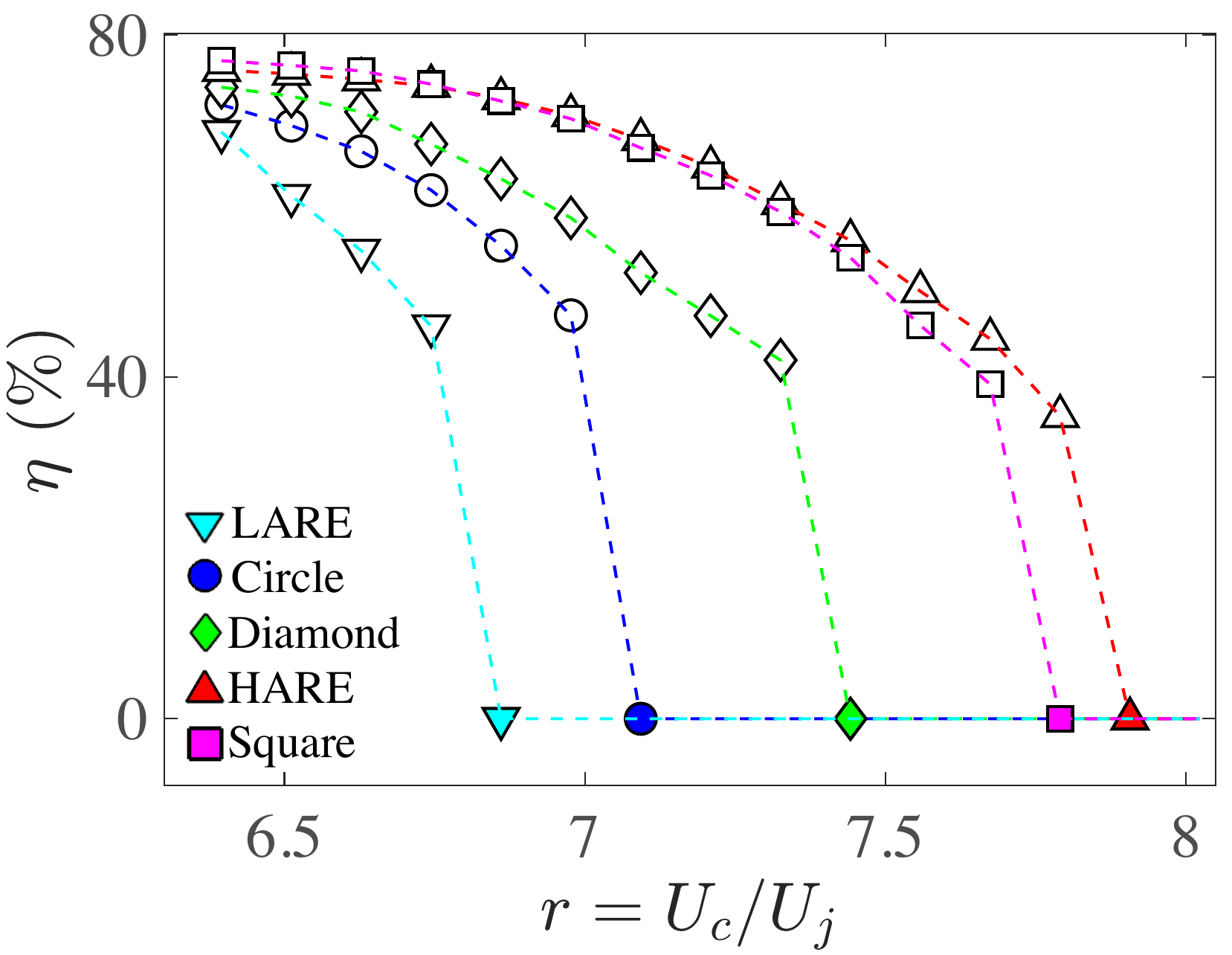}
    \caption{Combustion efficiency as a function of velocity ratio for different nozzle shapes with fixed $U_j = 3.4$ m/s. Open symbols indicate velocity ratios where there is no blow-out, closed symbols indicate blow-out velocity ratios. LARE (cyan inverted triangle), circle (blue circle), diamond (green diamond), HARE (red triangle), square (magenta square).}
    \label{fig:blowout}
\end{figure}

Figure~\ref{fig:blowout} illustrates the variation of combustion efficiency as a function of the velocity ratio for different nozzle geometries at fixed jet velocity, revealing distinct differences in blowout resistance among the nozzle geometries. Kalghati \cite{kalghatgi1981blow} empirically derived a universal non-dimensional stability curve for the blow-out stability of diffusion flames in a crosswind for different fuels. Using Kalghatgi's universal non-dimensional stability curve, the blow-out crosswind velocity for a round-pipe methane flare with a jet diameter $D = 7.62$ cm and jet velocity $U_j = 3.4$ m/s is $\approx$ 22 m/s. 

In the LES cases, all the nozzles were initialized with a crosswind velocity of 22 m/s and were simulated for 450$D/U_j$ (10 seconds). If the flame did not blow-out ($\eta = 0$), the crosswind velocity was increased by 0.4 m/s. This was repeated until all the nozzles experienced blow-out and $\eta = 0$.
The LARE nozzle exhibits the earliest onset of blowout, with efficiency plummeting at lower $r$ compared with the other designs. The circular nozzle follows, also showing a rapid decline at moderately high velocity ratios. In contrast, the diamond, HARE, and the square nozzle sustain higher combustion efficiencies over a broader range of crosswind velocities, with the square and HARE nozzles exhibiting the highest resilience to blowout. 

Notably, the LES-predicted blow-out velocity for the circular nozzle was $\approx$ 24 m/s, which is within 10\% of the experimental value ($\approx 22$ m/s) \cite{kalghatgi1981blow}.  This agreement demonstrates important validation of the model and that the modeling approach is capable of predicting the velocity of blow-out
(extinction) at the conditions studied.

\begin{figure}[h!]
    \centering
    \includegraphics[width=6in]{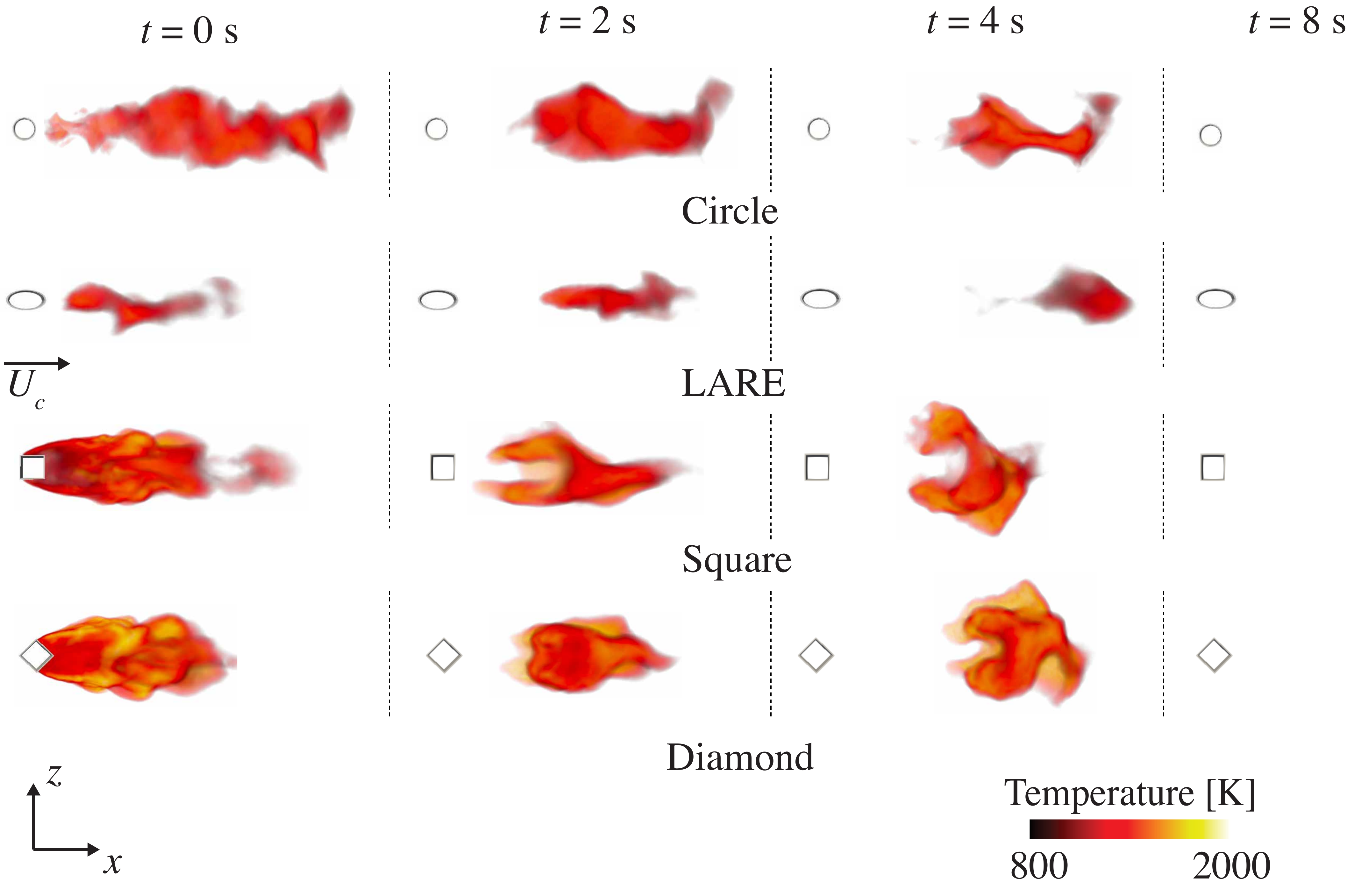}
    \caption{Temporal evolution of flame shapes for $U_j = 3.4$ m/s for the various nozzle geometries as conditions progress to blow-out  The instantaneous flame profiles are visualized as volume renderings of the temperature field for $T>800$ K. The time series of snapshots progresses from left to right. } 
    \label{fig:Blowshapes}
\end{figure}

Figure~\ref{fig:Blowshapes} presents instantaneous snapshots of volume rendered temperature fields for the circle, LARE, square and diamond nozzles during the flame blow-out process. These visualizations reveal distinct blow-out dynamics associated with each nozzle geometry. For the square nozzle, the flame initially remains anchored near the nozzle exit, indicating robust stabilization afforded by strong recirculation and corner-induced turbulence. As the blow-out event progresses, the flame rapidly detaches and is extinguished, reflecting an abrupt transition from a well-attached to a fully blown-out state.

In contrast, the circle and LARE nozzles exhibit different behavior. The flame becomes lifted even before the blowout velocity is reached. The lift-off height gradually increases with an increase in crosswind velocity, and the flame base moves downstream as the stabilizing recirculation zone weakens under the influence of increased crosswind. The lifted configuration persists for several instances, with the flame core (regions of $T>800$ K) residing further from the nozzle, until the flame ultimately detaches and is blown off completely. The extended residence time of the lifted flame downstream of these nozzles highlights their lower resistance to aerodynamic strain and diminished ability to promote recirculation to sustain a stable flame base.  Liu et al. \cite{liu2020influence} also reported similar flame attachment behavior, where they observed that streamlined nozzle geometries that did not promote a strong recirculation zone demonstrated a lifted flame that progressively moved downstream with an increase in crosswind velocities, whereas the nozzle geometries that promoted a strong recirculation zone resulted in an attached flame base for comparatively higher crosswind velocities and were resistant to blow-out.

\begin{figure}[h!]
    \centering
    \includegraphics[width=3.3in]{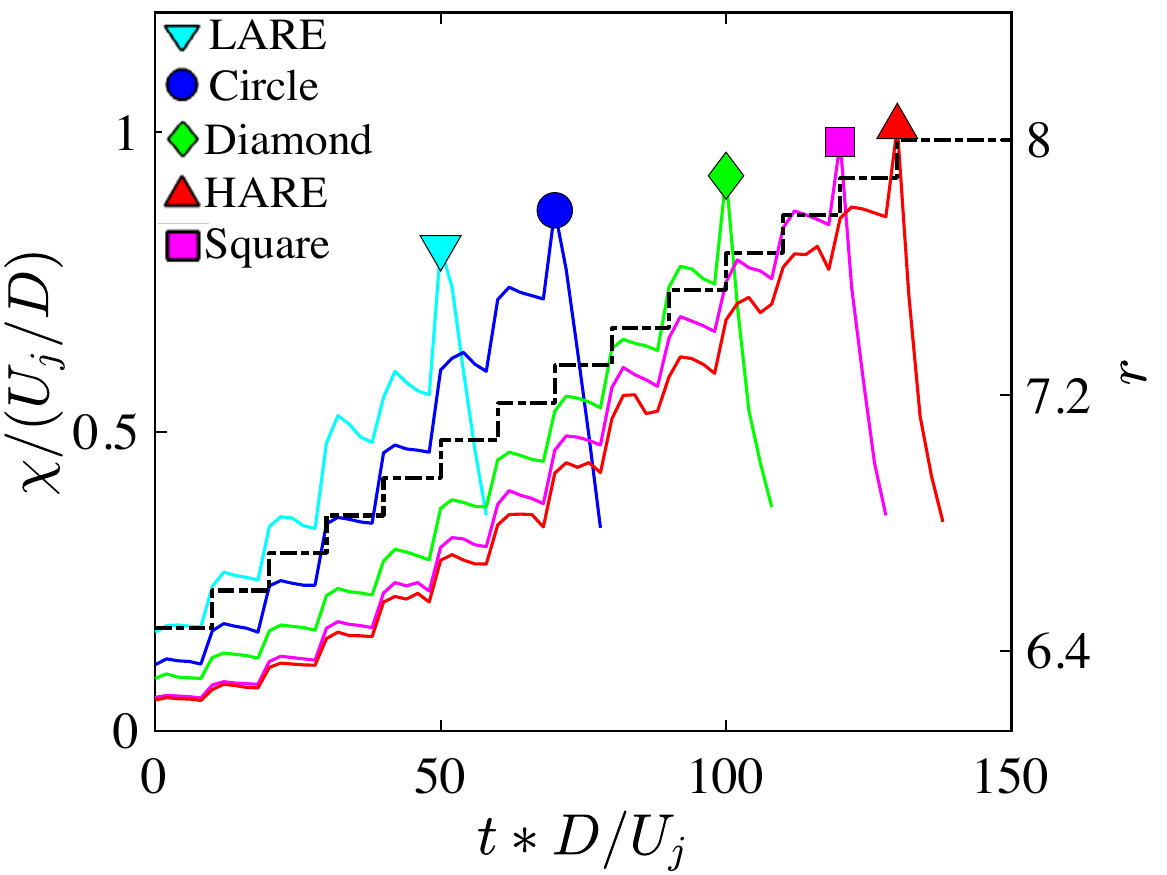}
    \caption{Temporal variation of the local scalar dissipation rate (left ordinate) at the flare tip and velocity ratio (right ordinate). The instantaneous scalar dissipation rate is calculated as a volume average within a cylindrical region of diameter $D$ and height $0.2D$, positioned immediately above the nozzle exit. The filled symbols represent the local quenching scalar dissipation rate ($\chi_q$). The velocity ratio for each simulation is provided for reference (black dotted line). LARE (cyan inverted triangle), circle (blue circle), diamond (green diamond), HARE (red triangle), square (magenta square).}
    \label{fig:strainrate_time}
\end{figure}

Figure~\ref{fig:strainrate_time} presents the temporal evolution of the local mean scalar dissipation rate, $\chi/(U_j/D)$,  at the exit of the flare for each nozzle geometry. The crosswind velocity ratio, $r = U_c/U_j$ (shown as a black dotted line, right axis), is incrementally increased in steps. The colored lines correspond to the strain rate evolution for each nozzle, with symbols marking the point at which the flame undergoes blow-out, corresponding to the local quenching scalar dissipation rate $\chi_q$, which represents the critical threshold of mixing above which the flame cannot be sustained. $\chi_q$ is commonly used to characterize flame extinction, as it signifies the limit at which turbulent mixing overwhelms the chemical reactions needed to maintain combustion \cite{sardi2000extinction,peters2001turbulent,poinsot1991quenching}.

The results demonstrate that as the crosswind is incrementally intensified, the near-nozzle strain rate exhibits a stair-step-like rise for all nozzles. The LARE and circle nozzles reach their quenching strain rates at earlier times and lower crosswind velocities, indicative of their increased susceptibility to blow-out. In contrast, the diamond, square, and HARE nozzles sustain lower strain rates for longer durations before reaching their respective quenching points. The square and HARE nozzles, in particular, exhibit the highest tolerance to increasing crosswind strain, consistent with their enhanced resistance to blow-out observed in earlier figures.

These findings highlight the role of local scalar dissipation in dictating flame stability under crosswind conditions. Nozzle geometries that promote enhanced recirculation by reintroducing hot combustion products into the near nozzle reaction zone due to the corner vortices, such as square and HARE, effectively buffer the flame from rapid increases in local strain, enabling sustained combustion even under aggressive aerodynamic strain. Conversely, designs with weaker recirculation and less turbulence are prone to rapid increases in strain, leading to earlier flame extinction. Thus, the interplay between nozzle-induced flow dynamics and local strain rates is key to achieving flare designs with robust resistance to blow-out and high combustion efficiency.

The results of this work highlight several new aspects of flare performance by revealing how simple geometric modifications to nozzle shape can meaningfully influence near-field mixing, combustion efficiency, and resistance to blowout under realistic crosswind conditions. Specifically, our findings show that corner-induced recirculation and enhanced turbulence—achieved without added operational complexity—can passively promote more robust flame stability and mixing. This approach differs from previous studies, which have often considered these performance metrics in isolation or emphasized more complex engineered solutions. By systematically examining a range of canonical nozzle shapes using high-fidelity LES and the FPVA approach, this work provides a more unified framework for understanding how geometry-induced flow features connect with practical flare performance. As such, these insights complement and extend existing literature, offering practical guidance for improving non-assist flare designs within field-relevant constraints.
%%%%%%%%%%%%%%%%%%%%%%%%%%%%%%%%%%%%%%%%%%%%%%%%%%%%%%%%%%%%%%%%%%%%%%%%%%%%%%%%%%%

\section{Conclusion}\label{sec4}

This study investigates the influence of nozzle geometry on combustion efficiency, mixing, and blowout dynamics for non-assist methane flares subjected to crosswind, using high-fidelity large-eddy simulations with a flamelet progress variable approach. By systematically comparing canonical nozzle designs—including square, diamond, high-aspect ratio ellipse, low-aspect ratio ellipse, and circle--across a range of crosswind velocities and directions, we establish clear links between geometric features, flow physics, and combustion performance.

Nozzles incorporating sharp corners and edges (such as square and diamond cross-sectional shapes) generate strong recirculation zones and corner-induced turbulence at the base of the flame. These features enhance local mixing, maintain high-temperature reaction zones near the nozzle, and promote flame attachment even under elevated crosswind. In particular, the square and diamond nozzle shapes show improved combustion efficiency regardless of wind direction compared with the widely-used circular pipe cross-section. As a result of the sharp corners, the square and diamond nozzle geometries consistently achieve higher combustion efficiency (up to $5\%$ at crosswind velocity of 12 m/s and jet velocity of 3.4 m/s), greater resistance to flame liftoff, and significantly delayed blowout compared to smooth, axisymmetric nozzles (circle and LARE). The square nozzle, in particular, demonstrated the lowest combustion inefficiency (average $\eta\approx0.5\%$ compared to $\eta\approx2\%$ for the circle nozzle) and highest tolerance to crosswind-induced strain, remaining compliant with EPA regulatory standards (i.e., $\eta>96.5\%$ ) under the wide range of conditions ($0\leq U_c\leq 12 $ m/s).

A key insight emerging from this work is that the analysis of mixing parameters, vorticity, and scalar dissipation suggests corner-driven flow structures can promote more rapid homogenization of the fuel–air mixture and provide some resilience against aerodynamic strain. Conversely, the results indicate that circular and LARE nozzles, due to their smoother profiles, tend to exhibit weaker recirculation and more pronounced flame stretching, which can contribute to increased fuel slip and lower overall efficiency, particularly under stronger crosswind conditions.

The present study not only confirms and extends previous experimental observations but also offers quantitative, geometry-specific effects on performance metrics for realistic flare operating conditions. Importantly, a simple change in the nozzle geometry is demonstrated as a practical, passive strategy for enhancing both combustion efficiency and blowout resistance, with direct implications for emissions reduction and regulatory compliance in field applications. These insights provide a robust foundation for future experimental validation and the development of next-generation flare designs capable of sustaining efficient, clean combustion under challenging real-world operating conditions.

\section{Acknowledgement}
The information, data, or work presented herein was funded
in part by the Advanced Research Projects Agency-Energy (ARPA-E), U.S. Depart-
ment of Energy, under Award Number DE-AR0001534. The views and opinions of
authors expressed herein do not necessarily state or reflect those of the United States Government or any agency thereof. This research was supported in part through computational resources and services provided by Advanced Research Computing (ARC), a division of Information and Technology Services (ITS) at the University of Michigan, Ann Arbor.

\section{Funding}
The information, data, or work presented herein was funded in part by the Advanced Research Projects Agency-Energy (ARPA-E), U.S. Department of Energy, under Award Number DE-AR0001534. The views and opinions of authors expressed herein do not necessarily state or reflect those of the United States Government or any agency thereof.

\subsection*{Notes}
The authors declare no competing financial interests.

\bibliography{acs-achemso}

% \begin{tocentry}

% \vspace*{0.32cm} % Adjust vertical centering as needed
% \includegraphics[width=8.25cm,height=4.45cm,keepaspectratio]{For_Table_of_Contents_Only.eps}

% % Optional: Add a single-line title or label in Helvetica 8pt, no section titles or multi-line text
% % \vspace{0.2cm}
% % {\sffamily\fontsize{8}{9.6}\selectfont Your short TOC graphic label (optional)}

% \end{tocentry}

\end{document}